\documentclass[11pt,a4paper]{article}
\usepackage{authblk}
\usepackage{lmodern}
\usepackage{jheppub}
\usepackage{amsmath}
\usepackage{mathtools}
\usepackage{tikz}
\usepackage{empheq}
\usepackage{enumitem}
\usepackage{graphics}
\usepackage{wrapfig}
\usepackage{caption}
\usepackage{natbib}
\usepackage{xcolor}
\usepackage{framed}
\usepackage{array}
\usepackage{hyperref}
\usepackage{dsfont}
\definecolor{shadecolor}{gray}{0.925}

\usepackage[cbgreek]{textgreek}

\def\sideremark#1{\ifvmode\leavevmode\fi\vadjust{\vbox to0pt{\vss
 \hbox to 0pt{\hskip\hsize\hskip1em
 \vbox{\hsize3cm\tiny\raggedright\pretolerance10000
 \noindent #1\hfill}\hss}\vbox to8pt{\vfil}\vss}}}%

\newcommand{\bi}{\begin{itemize}}
\newcommand{\ei}{\end{itemize}}
\newcommand{\bea}{\begin{align}}
\newcommand{\eea}{\end{align}}
\newcommand{\be}{\begin{equation}}
\newcommand{\ee}{\end{equation}}

\makeatletter
\renewcommand*\env@matrix[1][\arraystretch]{%
  \edef\arraystretch{#1}%
  \hskip -\arraycolsep
  \let\@ifnextchar\new@ifnextchar
  \array{*\c@MaxMatrixCols c}}
\makeatother
\author[\ensuremath{a,b,c,d}]{Francesca PACIFICO}
\author[\ensuremath{a,c,d}]{\quad Charlotte SLEIGHT}
\author[\ensuremath{c,d,e}]{\quad Massimo TARONNA}

\affiliation[\ensuremath{a}]{Department of Mathematical Sciences, \\ Durham University, Durham, DH1 3LE, U.K.}

\affiliation[\ensuremath{b}]{Dipartimento di Matematica, Universit\`a di Bologna, \\ Piazza di Porta S. Donato 5, 40126 Bologna, Italy}

\affiliation[\ensuremath{c}]{Dipartimento di Fisica ``Ettore Pancini'', Universit\`a degli Studi di Napoli Federico II, \\Monte S. Angelo, Via Cintia, 80126 Napoli, Italy}

\affiliation[\ensuremath{d}]{INFN, Sezione di Napoli, Monte S. Angelo, Via Cintia, 80126 Napoli, Italy}

\affiliation[\ensuremath{e}]{Scuola Superiore Meridionale, Universit\`a degli Studi di Napoli Federico II,\\ Largo San Marcellino 10, 80138 Napoli, Italy}

\emailAdd{francesca.pacifico@unina.it, charlotte.sleight@na.infn.it, massimo.taronna@unina.it}

\title{\centering \Huge Conformal Partial Wave Expansion \\ of Celestial Correlators}

\abstract{A novel definition of holographic correlation functions on the celestial sphere of Minkowski space was recently introduced in \cite{Sleight:2023ojm} as the extrapolation of bulk time-ordered correlation functions to the celestial sphere. In this work, focusing on theories of scalar fields in $(d+2)$-dimensional Minkowski space, we show that in perturbation theory such celestial correlation functions admit a conformal partial wave expansion with meromorphic spectral density, and hence also an expansion into conformal blocks. This is achieved in the hyperbolic slicing of Minkowski space by extending the harmonic function (``spectral") decomposition of AdS bulk-to-bulk propagators to the Minkowski Feynman propagator. We study the conformal partial wave expansion of celestial correlators for four-point contact and tree-level exchange diagrams, and extract the contributions to their conformal block expansions in the direct channel. When all scalar fields are massless, the tree-level exchange diagram takes a remarkably simple form and is given by a  finite sum of conformal blocks (and, for $d=2$, their derivatives as well). We also discuss the conformal partial wave expansion at the non-perturbative level, where Lorentz unitarity manifests as positivity of the spectral density.}

\begin{document}

\begin{flushright}    
\texttt{}
\end{flushright}

\maketitle

\newpage

\section{Introduction}

The celestial holography programme \cite{Raclariu:2021zjz,Pasterski:2021rjz,McLoughlin:2022ljp,Pasterski:2021raf} aims to apply the holographic principle to asymptotically flat space-times, proposing that quantum gravity can be encoded in a co-dimension two dual theory living on the celestial sphere at null infinity. Lorentz transformations act on the celestial sphere like conformal transformations and the holographic observables therefore take the form of conformal correlators on the celestial sphere. Indeed, through a certain change of basis scattering amplitudes in Minkowski space can be reorganised as correlation functions of conformal primaries on the celestial sphere \cite{Pasterski:2016qvg,Pasterski:2017kqt}, which we refer to as \emph{celestial amplitudes}. Many efforts have therefore been dedicated to studying the structure and symmetries of the corresponding celestial conformal field theory by translating insights from S-matrix theory. 

\vskip 4pt
It is natural to ask if any insights might also be gained from the relatively well understood example of holography in anti-de Sitter (AdS) space. In the absence of a standard notion of an S-matrix in AdS space, the AdS-CFT correspondence \cite{Maldacena:1997re,Gubser:1998bc,Witten:1998qj} proposes that holographic observables in AdS space are given by extrapolation of bulk correlation functions to the co-dimension 1 boundary at spatial infinity, where they take the form of correlation functions of conformal primaries in Minkowski space. Various works have aimed to draw lessons for celestial amplitudes from the AdS-CFT correspondence, such as by considering hyperbolic foliation \eqref{hypf} of Minkowski space \cite{deBoer:2003vf,Cheung:2016iub,Casali:2022fro,Iacobacci:2022yjo,Melton:2023bjw,Melton:2024jyq} and the flat space limit \cite{deGioia:2022fcn,deGioia:2023cbd}.

\vskip 4pt
Inspired by the extrapolate definition of holographic observables in AdS space, it was recently proposed in \cite{Sleight:2023ojm} that a definition of holographic observables on the celestial sphere could be given by extrapolating bulk correlation functions to the celestial sphere. Indeed, similar to Minkowski space, massive particles in AdS do not reach the boundary at infinity though such extrapolate holographic observables, being off-shell, place massive and massless fields on the same footing. The recipe to extrapolate correlation functions in Minkowski space to the celestial sphere naturally arises by considering a hyperbolic foliation:
\begin{equation}\label{hypf}
   X = R {\hat X}, \qquad X^2 = \sigma\, R^2,
\end{equation}
with radial coordinate $R>0$ and hyperbolic coordinates ${\hat X}^2 = \sigma$ where $\sigma \in \left\{+1,-1\right\}$. In this picture celestial sphere arises as the projective cone of light rays
\begin{equation}
Q^2=0, \qquad Q \equiv \lambda Q, \qquad \lambda \in \mathbb{R}^+,
\end{equation}
which is the conformal boundary of each hyperbolic slice in the foliation \eqref{hypf}. The extrapolation of (time-ordered) correlation functions of bulk fields $\phi_i\left(X_i\right)$ to the celestial sphere is then implemented by a Mellin transform in the radial direction followed by the boundary limit in the hyperbolic directions \cite{Sleight:2023ojm}:
\begin{align}\label{ccdefnint}
    \left\langle O_{\Delta_1}(Q_1)\ldots O_{\Delta_n}(Q_n)\right\rangle =\prod_i \lim_{{\hat X}_i\to Q_i}\,\int^\infty_0 \frac{{\rm d}R_i}{R_i}\,R_i^{\Delta_i}\left\langle\phi_1(R_1\hat{X}_1)\ldots \phi_n(R_n\hat{X}_n)\right\rangle.
\end{align}  
To distinguish them from celestial amplitudes described above, we refer to the extrapolation \eqref{ccdefnint} of Minkowski correlators to the celestial sphere as \emph{celestial correlators}. Feynman rules for the perturbative computation of such celestial correlators were detailed in \cite{Iacobacci:2024nhw} for theories of scalar fields and simply follow from the Feynman rules for (time-ordered) correlation functions in Minkowski space. Through the hyperbolic foliation \eqref{hypf}, celestial correlators \eqref{ccdefnint} share some similarities with holographic correlators in AdS space. In fact, it turns out that they can be perturbatively re-written in terms of corresponding Witten diagrams in a co-dimension 1 Euclidean AdS space \cite{Iacobacci:2024nhw}, which one might then use to import techniques and results from the relatively well understood AdS case.

\vskip 4pt
An instrumental tool that has led to several key insights in the study of boundary correlation functions in AdS is the conformal partial wave expansion (CPWE). This is the expansion of single-valued conformal correlation functions over an orthogonal basis of single-valued Eigenfunctions ${\cal F}_{\nu,J}$ of the Casimir invariants of the conformal group \cite{Dobrev:1975ru,Dobrev:1977qv,Mack:2009mi,Costa:2012cb,Caron-Huot:2017vep}. For a conformal four-point function of primary operators $O_{\Delta_i}$ this takes the form
\begin{equation}\label{cpweint}
    \langle {\cal O}_{\Delta_1}\left(Q_1\right){\cal O}_{\Delta_2}\left(Q_2\right){\cal O}_{\Delta_3}\left(Q_3\right){\cal O}_{\Delta_4}\left(Q_4\right) \rangle = \sum\limits_{J}\int^{\infty}_{-\infty}\frac{{\rm d}\nu}{2\pi}\,\rho_{J}\left(\nu\right){\cal F}_{\nu,J}\left(Q_1,Q_2,Q_3,Q_4\right),
\end{equation}
with spectral density $\rho_{J}\left(\nu\right)$. For boundary correlators in AdS, the spectral density $\rho_{J}\left(\nu\right)$ is a meromorphic function of $\nu$, which implies that they also admit a convergent expansion in terms of a discrete sum of conformal blocks. The fact that celestial correlators \eqref{ccdefnint} can be re-cast in terms of corresponding Witten diagrams in EAdS suggests that, at least perturbatively, they also admit a conformal partial wave expansion \eqref{cpweint} with meromorphic spectral density and hence, an expansion in conformal blocks as well. In this paper we will show that this is indeed the case, focusing for simplicity on theories of scalar fields.\footnote{The conformal partial wave and conformal block expansion of celestial amplitudes \cite{Pasterski:2016qvg,Pasterski:2017kqt} (as opposed to celestial correlators \eqref{ccdefnint}) has been considered in various works \cite{Lam:2017ofc,Nandan:2019jas,Atanasov:2021cje,Melton:2021kkz,Iacobacci:2022yjo,Chang:2023ttm,Fan:2023lky,Liu:2024vmx,Kulp:2024scx}. \label{foo::1}}

\vskip 4pt
A useful tool to determine the conformal partial wave expansion of Witten diagrams in anti-de Sitter space is the decomposition of the bulk-to-bulk propagator in terms of an orthogonal basis of Eigenfunctions $\Omega_{\nu,J}$ of the AdS Laplacian (see e.g. \cite{Moschella:2007zza,Penedones:2007ns,Costa:2014kfa}). In this work we extend the harmonic function decomposition to the Feynman propagator, which plays the role of the bulk-to-bulk propagator for the celestial correlators \eqref{ccdefnint}. In particular, we show that the Feynman propagator $G^{(m)}_T$ for a scalar field of mass $m$ in Minkowski space can be decomposed in the different regions of the hyperbolic slicing \eqref{hypf} in terms of Euclidean AdS (EAdS) harmonic functions (via analytic continuation), taking the form:
\begin{shaded}
\noindent \emph{EAdS harmonic function decomposition of the Feynman propagator.}
 \begin{align} \nonumber
\hspace*{-0.75cm}G^{(m)}_T(X_{{\cal A}_+},Y_{{\cal A}_+})&=\int_{-\infty}^{+\infty}\frac{{\rm d}\nu}{2\pi}\, {}^{(m)}a^{{\cal A}_+,{\cal A}_+}_\nu(R_1,R_2)  \Omega_{\nu,0}({\hat X}_{{\cal A}_+},{\hat Y}_{{\cal A}_+}),\\ \nonumber
 \hspace*{-0.75cm}  G^{(m)}_T(X_{{\cal D}_+},Y_{{\cal D}_+})
   &\to  \int_{-\infty}^{+\infty}\frac{{\rm d}\nu}{2\pi}{}^{(m)}a^{{\cal D}_+,{\cal D}_+}_\nu(R_1,R_2)\Omega_{\nu,0}({\hat X}_{{\cal A}_+},{\hat Y}_{{\cal A}_+}),\\ 
 \hspace*{-0.75cm}  G^{(m)}_T(X_{{\cal D}_+},Y_{{\cal A}_+})&\to  \int_{-\infty}^{+\infty}\frac{{\rm d}\nu}{2\pi}\,{}^{(m)}a^{{\cal D}_+,{\cal A}_+}_\nu(R_1,R_2)\Omega_{\nu,0}({\hat X}_{{\cal A}_+},{\hat Y}_{{\cal A}_+}), \label{bubuharm0}
\end{align}   
\end{shaded}
\noindent where ${\cal A}_+$ and ${\cal D}_+$ are, respectively, the EAdS and dS regions \eqref{AHYP} and \eqref{DHYP} of Minkowski space that contribute to the celestial correlators \eqref{ccdefnint}. The arrows denote analytic continuation from the de Sitter region ${\cal D}_+$ to the EAdS region ${\cal A}_+$. The spectral functions ${}^{(m)}a^{\bullet,\bullet}_\nu(R_1,R_2)$ in each region encode the dependence on the radial direction and the particle mass and are given explicitly in section \ref{subsec::adsfeyn}. The result \eqref{bubuharm0} essentially follows from the decomposition \cite{Iacobacci:2024nhw} of the Feynman propagator in terms of bulk-to-bulk propagators in EAdS carrying principal series representations of the Lorentz group.

\vskip 4pt
Using the harmonic function decomposition \eqref{bubuharm0} of the Feynman propagator, one can determine the conformal partial wave expansion of perturbative celestial correlators \eqref{ccdefnint} along the same lines as for Witten diagrams in EAdS, upon evaluating the radial integrals.\footnote{Since \eqref{bubuharm0} concerns internal legs of Feynman diagrams, it should be applicable to celestial amplitudes \cite{Pasterski:2016qvg,Pasterski:2017kqt} as well. It would be interesting to make contact with the works referenced in footnote \ref{foo::1} via the spectral decomposition \eqref{bubuharm0} of the Feynman propagator.} Meromorphicity in the spectral parameter $\nu$ for perturbative celestial correlators then follows from the fact that it holds in EAdS perturbation theory and the structure of the spectral functions ${}^{(m)}a^{\bullet,\bullet}_\nu(R_1,R_2)$. In this work we work this out explicitly for tree-level exchange diagrams in scalar field theories and study the contributions to the corresponding conformal block expansion in the direct channel. This is carried out in detail for four-point exchange diagrams with external massless scalar fields, which are simpler than their massive counterparts owing to the simplicity of the Feynman propagator in this case. 

\vskip 4pt
Like for four-point Witten diagrams in AdS space \cite{Penedones:2010ue,ElShowk:2011ag}, the conformal block expansion of four-point contact diagrams and tree-level exchange diagrams on the celestial sphere generally contain contributions from two infinite families of composite ``double-trace" operators which have the schematic form
\begin{equation}
    \left[{\cal O}_{\Delta_1}{\cal O}_{\Delta_2}\right]_{n}\sim {\cal O}_{\Delta_1} (\partial^ 2)^ n {\cal O}_{\Delta_2}+\ldots, \qquad \left[{\cal O}_{\Delta_3}{\cal O}_{\Delta_4}\right]_{n}\sim {\cal O}_{\Delta_3} (\partial^ 2)^ n {\cal O}_{\Delta_4}+\ldots\,,
\end{equation}
and scaling dimensions:
\begin{align}
    &\Delta_1+\Delta_2+2n, \qquad n =0, 1, 2, \ldots, \\
    &\Delta_3+\Delta_4+2n, \qquad n =0, 1, 2, \ldots.
\end{align}
Particle exchanges are encoded by further contributions to the conformal block expansion. For celestial correlators of massless scalars, when the exchanged particle is massive these are encoded by two infinite families of \emph{double-trace-like} contributions with scaling dimensions:
\begin{align}\label{radialDT0}
    &\Delta_1+\Delta_2+(2-d)+2n, \qquad n = 0, 1, 2, \ldots,\\
     &\Delta_3+\Delta_4+(2-d)+2n, \qquad n = 0, 1, 2, \ldots,
\end{align}
which encode the massive exchanged single-particle state in $(d+2)$-dimensional Minkowski space. These contributions were also recently obtained from the Mellin amplitude representation \cite{Pacifico:2024dyo} of celestial correlation functions and here we show the equivalence between the latter and the conformal partial wave expansion of celestial correlation functions. The harmonic function decomposition \eqref{bubuharm0} provides further insights on the structure of celestial correlators. For example, it makes manifest that the single-particle contributions \eqref{radialDT0} originate from the integral over the radial direction, as well as meromorphicity of the spectral density in the conformal partial wave expansion for perturbative celestial correlators.

\vskip 4pt
For celestial correlators involving only massless scalars, there is an additional linear constraint on the scaling dimensions originating from the power-law dependence of the massless scalar propagators on the radial direction. For exchange diagrams, this leads to drastic simplifications in the conformal block expansion and the final result for integer dimensions $d$ turns out to be a \emph{finite} sum of leading (i.e. $n=0$) double-trace and double-trace-like contributions \eqref{radialDT0}:
\begin{shaded}
\noindent \emph{Tree-level exchange for massless scalars (integer $d$).}
\begin{multline}
 \langle {\cal O}_{\Delta_1}\left(Q_1\right){\cal O}_{\Delta_2}\left(Q_2\right){\cal O}_{\Delta_3}\left(Q_3\right){\cal O}_{\Delta_4}\left(Q_4\right) \rangle= 2\pi i\,\delta\left(\frac{4-3d+\sum_i\Delta_i}{2}\right) \,  \\ \times 2 \sin \left(\frac{\pi  d}{2}\right) \Gamma \left(\frac{d-2}{2}\right)^2 \Gamma \left(-\tfrac{d}{2}+\Delta_1+1\right) \Gamma \left(-\tfrac{d}{2}+\Delta_2+1\right) \Gamma (d-\Delta_3-1) \Gamma (d-\Delta_4-1)\\
   \times\frac{\sin (\pi  (d-\Delta_1-\Delta_2)) \Gamma \left(\tfrac{3 d}{2}-\Delta_1-\Delta_2-2\right) \Gamma (2 d-\Delta_1-\Delta_2-3)}{-d+\Delta_1+\Delta_2+1}\ {\cal G}^{12,34}_{\Delta_1+\Delta_2+2-d,0}\left(Q_i\right)\\
    +(\Delta_1,\ \Delta_2\ \leftrightarrow\ \Delta_3,\ \Delta_4),
\end{multline}
\end{shaded}
\noindent where ${\cal G}^{12,34}_{\Delta,J}$ denotes a conformal block associated to a primary operator of scaling dimension $\Delta$ and spin-$J$ in the (12)(34) operator product expansion (OPE) channel. Interestingly, this vanishes for even $d > 2$ if the corresponding conformal block does not diverge (and up to contact terms).  For $d=2$ instead, the zero is canceled by a pole and the above expression becomes a sum of the two conformal blocks and their derivatives, given in equation \eqref{exchd2}.\footnote{A similar mechanism occurs for boundary correlators in de Sitter space as well \cite{Sleight:2019mgd,Sleight:2019hfp,Sleight:2020obc,Sleight:2021plv}, where zeros generated by the bulk time integration can cancel \cite{Sleight:2019mgd} potential divergent contributions which are usually related to IR divergences.} This simplification with respect to the exchange of a massive scalar field is owing to the one-to-one map between massless fields in Minkowski space and massive fields on the co-dimension 1 hyperboloid \cite{Fronsdal:1978vb}, which therefore extends to the exchange of massless spinning fields in Minkowski space as well \cite{toappear}.

\vskip 4pt
Given that perturbative celestial correlators admit a conformal partial wave expansion \eqref{cpweint} with meromorphic spectral density, it is tempting to explore the implications should it hold non-perturbatively as well. In this spirit, at the end of this work we propose a non-perturbative conformal bootstrap of celestial correlation functions \eqref{ccdefnint}, where unitarity with respect to the Lorentz group manifests as positivity of the spectral density.

\newpage 
The outline of this paper is as follows:

\begin{itemize}
    \item In section \ref{sec::CCF} we review the prescription for correlation functions \eqref{ccdefnint} on the celestial sphere of $(d+2)$-dimensional Minkowski space and their Feynman rules. We review the radial reduction of the latter onto the extended unit hyperboloid in Minkowski space and their relation to propagators in Euclidean AdS$_{d+1}$, which in the de Sitter region is by analytic continuation.
    \item In section \ref{sec::CPWE} we begin by reviewing the harmonic function decomposition of bulk-to-bulk propagators in EAdS and how it is used to obtain the conformal partial wave and conformal block expansion of Witten diagrams. In section \ref{subsec::adsfeyn} we use the hyperbolic slicing \eqref{hypf} to extend the harmonic function decomposition to the Minkowski Feynman propagator for a scalar field, which follows its decomposition \cite{Iacobacci:2024nhw} in terms of EAdS bulk-to-bulk propagators carrying Principal Series representations of the Lorentz group. In section \ref{subsecc::TLexch} this is applied to obtain the conformal partial wave expansion of four-point celestial correlators \eqref{ccdefnint} generated by a tree-level exchange diagram in scalar field theory and discuss the corresponding contributions to the conformal block expansion. We point out some subtleties with the choice of contour in the conformal partial wave expansion for perturbative celestial correlators, which do not appear to arise in standard AdS Witten diagram computations.
    \item In section \ref{sec::MLcase} we consider celestial correlators \eqref{ccdefnint} for massless scalar fields. Celestial correlators involving massless scalar fields are simpler than their massive counterparts owing to the power-law dependence of their propagators on the radial direction. This, in turn, leads to linear constraints on the scaling dimensions in this case. In section \ref{subsec::exchdiagramecbw} we consider in detail the conformal partial wave expansion of four-point celestial correlators with massless external scalar fields that are generated by the tree-level exchange of a massive scalar field. We also extract the contributions to the corresponding conformal block expansion. In section \ref{subsec::mlexchsc} we consider the special case that the exchanged scalar field is also massless, which presents some subtleties with the contour of the spectral integral owing to the linear relations among scaling dimensions that emerge in this case. As as result, the exchange diagram in this case reduces to a finite sum of conformal blocks (and, for $d=2$, also their derivatives).
    \item In section \ref{sec::outlook} we conclude by proposing a non-perturbative bootstrap of celestial correlators \eqref{ccdefnint}, assuming that the conformal partial wave expansion \eqref{cpweint} with meromorphic spectral density $\rho_J\left(\nu\right)$ continues to hold at the non-perturbative level. Unitarity of the Euclidean conformal field theory on the celestial sphere corresponds to positivity of the spectral density $\rho_J\left(\nu\right) \geq 0$.
    \item Various technical details are relegated to the appendices. In appendix \ref{app::spectralintegral} we show the equivalence between the conformal partial wave decomposition of the four-point tree-level exchange diagram of scalar fields obtained in this work and its Mellin amplitude representation recently given in \cite{Pacifico:2024dyo}.
\end{itemize}

\newpage

\section{Celestial correlation functions}
\label{sec::CCF}

In this section we review the relevant aspects of the prescription \cite{Sleight:2023ojm} for celestial correlation functions in $(d+2)$-dimensional Minkowski space, including their Feynman rules and their reformulation in terms of Witten diagrams in EAdS$_{d+1}$ \cite{Iacobacci:2024nhw}.

\vskip 4pt
Celestial correlation functions were defined in \cite{Sleight:2023ojm}  as the extrapolation of time-ordered correlation functions in Minkowski space to the celestial sphere, in analogy to the extrapolate definition of holographic correlation functions in (A)dS. This is implemented in the hyperbolic slicing \eqref{hypf} of Minkowski space by a Mellin transform in the radial direction and a boundary limit ${\hat X}_i\to Q_i$ in the hyperbolic directions:
\begin{align}\label{ccdefn}
    \left\langle\mathcal{O}_{\Delta_1}(Q_1)\ldots \mathcal{O}_{\Delta_n}(Q_n)\right\rangle =\prod_i \lim_{{\hat X}_i\to Q_i}\,\int^\infty_0 \frac{{\rm d}R_i}{R_i}\,R_i^{\Delta_i}\left\langle\phi_1(R_1\hat{X}_1)\ldots \phi_n(R_n\hat{X}_n)\right\rangle.
\end{align}
The Feynman rules for the perturbative computation of such celestial correlators then follow from the standard Feynman rules for time-ordered Minkowski correlation functions upon extrapolating the external points to the celestial sphere according to \eqref{ccdefn}. These we given for scalar fields in \cite{Iacobacci:2024nhw}, which we review in the following. The bulk-to-bulk propagator is simply the Feynman propagator, which for a scalar field of mass $m$ reads,
\begin{equation}
    G^{(m)}_{T}\left(X,Y\right)=\frac{1}{2 \pi^{\frac{d+2}{2}}}\left(\frac{m}{2}\right)^{\frac{d}{2}} \frac{1}{\left[\left(X-Y\right)^2+i\epsilon\right]^{\frac{d}{4}}}K_{\frac{d}{2}}\left(m\left[\left(X-Y\right)^2+i\epsilon\right]^{\frac{1}{2}}\right),
\end{equation}
where $K_{\alpha}(z)$ is the modified Bessel function of the second kind. External (boundary) points on the celestial sphere are connected to bulk points via the \emph{celestial bulk-to-boundary propagator} \cite{Sleight:2023ojm}:
\begin{align}\label{celestialbubo}
    G^{\left(m\right)}_{\Delta}\left(X,Q\right) &=\lim_{\hat{Y}\to Q} \int^\infty_0 \frac{{\rm d}R}{R} R^\Delta\,G^{\left(m\right)}_T\left(X,R\hat{Y}\right).
\end{align}
 The extrapolation of both bulk points to the celestial sphere gives the free theory celestial two-point function:
\begin{align}\nonumber
    G^{(m)}_{\Delta_1 \Delta_2}(Q_1,Q_2)&=\lim_{{\hat X}_2\to Q_2,\, {\hat X}_1\to Q_1}\int^{\infty}_0\frac{{\rm d}R_2}{R_2}R^{\Delta_2}_2\int^{\infty}_0\frac{{\rm d}R_1}{R_1}R^{\Delta_1}_1 G^{\left(m\right)}_T\left(R_2 {\hat X}_2,R_1\hat{X}_1\right),\\
    &=\frac{C^{\text{flat}}_{\Delta_1}}{(-2Q_1\cdot Q_2+i\epsilon)^{\Delta_1}}(2\pi i)\delta(\Delta_1-\Delta_2)\,, \label{bdry2pt}
\end{align}
with normalisation 
\begin{equation}\label{celest2ptnorm}
   C^{\text{flat}}_{\Delta}= \left(\frac{m}2\right)^{d-2\Delta}\frac{1}{4\pi^{\frac{d+2}2}}\,\Gamma(\Delta)\Gamma(\Delta-\tfrac{d}2).
\end{equation}

\vskip 4pt
\paragraph{Radial reduction.} It is instructive to consider a radial reduction of the Feynman rules onto the extended unit hyperboloid ${\hat X}$, where irreducible representations of the hyperbolic isometry group i.e. the Lorentz group $SO(1,d+1)$ are labeled by a scaling dimension $\Delta$ and spin $J$. The celestial bulk-to-boundary propagator \eqref{celestialbubo} for a scalar field ($J=0$) of mass $m$ takes the form 
    \begin{align}\label{cbuboexpl}
    G^{\left(m\right)}_{\Delta}\left(X,Q\right)=\mathcal{K}^{(m)}_{i\left(\frac{d}{2}-\Delta\right)}\left(\sqrt{X^2+i\epsilon}\right)\,G_\Delta({\hat X},Q),
\end{align}
where the dependence on the hyperbolic directions is given by 
  \begin{equation}
    G_\Delta({\hat X},Q) = C^{\text{dS}}_{\Delta} \frac{\left(\sqrt{X^2+i\epsilon}\right)^{\Delta}}{\left(-2 X \cdot Q+i\epsilon\right)^{\Delta}},
    \end{equation}
and radial kernel\footnote{For later purposes it will be useful to note the radial kernel \eqref{rkernel} takes the form of the corresponding momentum space bulk-to-boundary propagator in EAdS$_{1-d}$ (\emph{c.f.} \cite{Freedman:1998tz}), where the mass $m$ takes the role of the modulus of the boundary momentum.}
    \begin{equation}
    \label{rkernel}
    \mathcal{K}^{(m)}_{i\left(\frac{d}{2}-\Delta\right)}\left(R\right)=\left(\frac{m}{2}\right)^{\tfrac{d}{2}-\Delta}\frac{2 R^{-\frac{d}{2}}}{\Gamma(\tfrac{d}{2}-\Delta)}\,K_{\Delta-\tfrac{d}{2}}(m R)\,.
\end{equation}  
The radial reduction of the Feynman propagator is \cite{Iacobacci:2024nhw}:
\begin{align}\label{spectralrepcelesybubu}
G^{(m)}_T(X,Y)=\frac{1}{2}\int_{-\infty}^{+\infty}\frac{{\rm d}\nu}{2\pi}\,\rho^{\left(m\right)}_\nu(\sqrt{X^2+i\epsilon},\sqrt{Y^2+i\epsilon}\,)G_{\frac{d}{2}+i\nu}(\sigma_\epsilon)\,,
\end{align}
where 
\begin{align}\label{rho}
\rho^{\left(m\right)}_\nu\left(R_1,R_2\right)&=\mathcal{K}_\nu^{(m)}(R_1)\mathcal{K}_{-\nu}^{(m)}(R_2\,),\\
G_{\Delta}\left(\sigma\right) &=\frac{\Gamma(\Delta)\Gamma(d-\Delta)}{(4\pi)^{\frac{d+1}2}\Gamma(\frac{d+1}2)}{}_2{F}_1\left(\begin{matrix}\Delta,d-\Delta\\\frac{d+1}{2}\end{matrix};\sigma\right),
\end{align}
and 
\begin{equation}
\sigma_\epsilon=\frac{1}{2}\left(1-\frac{-\hat{X}\cdot\hat{Y}+i\epsilon}{\sqrt{\hat{X}^2+i\epsilon}\sqrt{\hat{Y}^2+i\epsilon}}\right).
\end{equation}
Throughout this work the integral over $\nu$ should be regarded as a Mellin-Barnes type integral where the integration contour is usually a deformation of the real axis such that poles encoded in $\Gamma$-function factors of the form $\Gamma\left(a+i\nu\right)$ are separated from those of the form $\Gamma\left(b-i\nu\right)$.

\vskip 4pt
Inside the light cone, the $\left(d+1\right)$-dimensional hyperbolic foliating surfaces \eqref{hypf} are Euclidean anti-de Sitter spaces, with disconnected regions:
\begin{subequations}\label{AHYP}
 \begin{align}
&{\cal A}_+: \quad X^2 <0, \quad X^0 >0, \\
&{\cal A}_-: \quad X^2 <0, \quad X^0 <0.
\end{align}   
\end{subequations}
Outside the light cone the foliating surfaces are de Sitter spaces, with expanding and contracting regions
\begin{subequations}\label{DHYP}
 \begin{align}
&{\cal D}_+: \quad X^2 > 0, \quad X^+ >0, \\
&{\cal D}_-: \quad X^2 > 0, \quad X^+ <0,
\end{align}   
\end{subequations}
where $X^+$ is the light-cone coordinate $X^+=X^0+X^1$. Interestingly, in the perturbative computation of celestial correlators \eqref{ccdefn} the contributions from regions ${\cal A}_-$ and ${\cal D}_-$ in the hyperbolic slicing precisely cancel at all orders in perturbation theory \cite{Iacobacci:2024nhw}. We henceforth only consider contributions from internal points in regions ${\cal A}_+$ and ${\cal D}_+$.

\vskip 4pt
In region ${\cal A}_+$ the Feynman rules decompose in terms of propagators for scalar fields in EAdS:
\begin{subequations}
\begin{align}\label{cbuboap}
G^{\left(m\right)}_{\Delta}(X_{{\cal A}_+ },Q)&=c_\Delta^{\text{dS-AdS}} \,\mathcal{K}^{(m)}_{i\left(\frac{d}{2}-\Delta\right)}( + e^{\frac{\pi i}{2}} R\,)\, i^ {+\Delta}K^{\text{AdS}}_{\Delta}\left(  s_{\text{AdS}}({\hat X}_{{\cal A}_+},Q)-i\epsilon\right),\\
G^{(m)}_T(X_{{\cal A}_+},Y_{{\cal A}_+})&=\int_{-\infty}^{+\infty}\frac{{\rm d}\nu}{2\pi}\,\rho^{\left(m\right)}_\nu(+ e^{\frac{\pi i}{2}}R_1,+ e^{\frac{\pi i}{2}}R_2) \\ & \hspace*{5cm} \times  e^{\left(\frac{d}{2}+i\nu\right)\pi i}c^{\text{dS-AdS}}_{\frac{d}{2}+i\nu}\, G^{\text{AdS}}_{\frac{d}{2}+i\nu}({\hat X}_{{\cal A}_+},{\hat Y}_{{\cal A}_+}), \nonumber 
\end{align}    
\end{subequations}
where the bulk-to-boundary and bulk-to-bulk propagators for a scalar field of mass $m^2_{\text{AdS}}=\Delta\left(\Delta-d\right)$ in EAdS$_{d+1}$ are, respectively,
\begin{align}\label{buboads}
    K^{\text{AdS}}_{\Delta}\left(s_{\text{AdS}}\right)= \frac{C^{\text{AdS}}_{\Delta}}{\left(-2 s_{\text{AdS}} \right)^{\Delta}}, \qquad s_{\text{AdS}}({\hat X}_{{\cal A}_+},Q) = {\hat X}_{{\cal A}_+} \cdot Q,
\end{align}
and 
\begin{align}\label{bubuads}
    G^{\text{AdS}}_{\Delta}\left(\sigma_{\text{AdS}}\right) &= C^{\text{AdS}}_{\Delta} \left(-4\sigma_{\text{AdS}}\right)^{-\Delta} {}_2F_1\left(\begin{matrix}\Delta,\Delta-\frac{d}{2}+\frac{1}{2}\\2\Delta-d+1\end{matrix},\frac{1}{\sigma_{\text{AdS}}}\right),\\
    \sigma_{\text{AdS}}({\hat X}_{{\cal A}_+},{\hat Y}_{{\cal A}_+})&=\frac{1}{2}\left(1+{\hat X}_{{\cal A}_+} \cdot {\hat Y}_{{\cal A}_+}\right),
\end{align}
with coefficient 
\begin{equation}
    C^{\text{AdS}}_{\Delta} = \frac{\Gamma\left(\Delta\right)}{2\pi^{\frac{d}{2}}\Gamma\left(\Delta+1-\frac{d}{2}\right)}.
\end{equation}

\vskip 4pt
In region ${\cal D}_+$, the Feynman rules decompose in terms of time-ordered propagators for scalar fields in the Bunch-Davies vacuum \cite{Allen:1985ux} of dS:
\begin{align}
G^{\left(m\right)}_{\Delta}(X_{{\cal D}_+ },Q)&= \,\mathcal{K}^{(m)}_{i\left(\frac{d}{2}-\Delta\right)}(R)\, K^{\text{AdS}}_{\Delta, T}\left(s_{\text{AdS}}({\hat X}_{{\cal D}_+},Q)-i\epsilon\right), \\
G^{(m)}_T(X_{{\cal D}_+},Y_{{\cal D}_+})&= \frac{1}{2} \int_{-\infty}^{+\infty}\frac{{\rm d}\nu}{2\pi}\,\rho_\nu^{(m)}(R_1,R_2)\,G^{\text{dS}}_{\frac{d}{2}+i\nu,\, T}({\hat X}_{{\cal D}_+},{\hat Y}_{{\cal D}_+}),
\end{align}
where the time-ordered bulk-to-boundary and bulk-to-bulk propagators for a scalar field of mass $m^2_{\text{dS}}=\Delta\left(d-\Delta\right)$ in dS$_{d+1}$ are, respectively,
\begin{align}
    K^{\text{dS}}_{\Delta,T}\left(s_{\text{dS}}\right)&= \frac{C^{\text{dS}}_{\Delta}}{\left(-2 s_{\text{dS}} + i \epsilon \right)^{\Delta}},  \qquad s_{\text{dS}}({\hat X}_{{\cal D}},Q) = {\hat X}_{{\cal D}} \cdot Q,\\
    G^{\text{dS}}_{\Delta, T}\left(\sigma_{\text{dS}}\right) &= G_{\Delta}\left(\sigma_{\text{dS}}-i\epsilon\right), \hspace*{1cm} \sigma_{\text{AdS}}({\hat X}_{{\cal D}},{\hat Y}_{{\cal D}})=\frac{1}{2}\left(1+{\hat X}_{{\cal D}} \cdot {\hat Y}_{{\cal D}}\right),
\end{align}
with coefficient 
\begin{equation}
   C^{\text{dS}}_{\Delta}= \frac{1}{4\pi^{\frac{d+2}2}}\,\Gamma(\Delta)\Gamma(\tfrac{d}2-\Delta).
\end{equation}

\vskip 4pt
\paragraph{Reformulation in terms of EAdS Witten diagrams.} The fact that dS and EAdS spaces are related under analytic continuation implies their propagators are related under analytic continuation as well \cite{Sleight:2019mgd,Sleight:2020obc,Sleight:2021plv}. In particular, time-ordered points in ${\cal D}_+$ are mapped to points in ${\cal A}_+$ with the same curvature radius according to \cite{Sleight:2019mgd}:
\begin{equation}\label{dsadscont}
{\hat X}_{{\cal D}_+} \to +i {\hat X}_{{\cal A}_+},
\end{equation}
under which we have \cite{Sleight:2020obc,Sleight:2021plv}:
\begin{align}
    G^{\text{dS}}_{\Delta, T}({\hat X}_{{\cal D}_+},{\hat Y}_{{\cal D}_+}) &\,\to\, c^{\text{dS-AdS}}_{\Delta_+}e^{-\Delta_+\pi i}G^{\text{AdS}}_{\Delta_+}({\hat X}_{{\cal A}_+},{\hat Y}_{{\cal A}_+})+\,\left(\Delta_+ \to \Delta_-\right),\\
    K^{\text{dS}}_{\Delta,\,T}({\hat X}_{{\cal D}_+},Q)&\to  c^{\text{dS-AdS}}_{\Delta}\,i^{- \Delta}\,K^{\text{AdS}}_{\Delta}(s_{\text{AdS}}({\hat X}_{{\cal A}_+},Q)-i\epsilon),
\end{align}
with a coefficient that accounts for the change in two-point coefficient from AdS to dS:
\begin{equation}\label{dS-AdScoeff}
   c^{\text{dS-AdS}}_{\Delta} = \frac{C^{\text{dS}}_{\Delta}}{C^{\text{AdS}}_{\Delta}}=\frac{1}{2}\csc\left(\tfrac{\pi}{2}\left(d-2\Delta\right)\right).
   \end{equation}
Through the hyperbolic slicing of Minkowski space, this implies that the Feynman rules for celestial correlation functions \eqref{ccdefn} can also be completely reformulated in terms of propagators for scalar fields in EAdS$_{d+1}$. This can be summarised as follows (for details see \cite{Iacobacci:2024nhw}): 
\begin{align} \nonumber
G^{(m)}_T(X_{{\cal A}_+},Y_{{\cal A}_+})&=\int_{-\infty}^{+\infty}\frac{{\rm d}\nu}{2\pi}\,\rho_\nu^{(m)}(+ e^{\frac{\pi i}{2}}R_1,+ e^{\frac{\pi i}{2}}R_2) e^{\left(\frac{d}{2}+i\nu\right)\pi i}c^{\text{dS-AdS}}_{\frac{d}{2}+i\nu} G^{\text{AdS}}_{\frac{d}{2}+i\nu}({\hat X}_{{\cal A}_+},{\hat Y}_{{\cal A}_+}),\\ \nonumber
   G^{(m)}_T(X_{{\cal D}_+},Y_{{\cal D}_+})
   &\to \int_{-\infty}^{+\infty}\frac{{\rm d}\nu}{2\pi}\,\rho_\nu^{(m)}(R_1,R_2)e^{-(\frac{d}{2}+i\nu)\pi i}c^{\text{dS-AdS}}_{\frac{d}{2}+i\nu}G^{\text{AdS}}_{\frac{d}{2}+i\nu}({\hat X}_{{\cal A}_+},{\hat Y}_{{\cal A}_+}),\\ 
   G^{(m)}_T(X_{{\cal D}_+},Y_{{\cal A}_+})&\to \int_{-\infty}^{+\infty}\frac{{\rm d}\nu}{2\pi}\,\rho_\nu^{(m)}(R_1,R_2e^ {\frac{\pi i}{2}})c^{\text{dS-AdS}}_{\frac{d}{2}+i\nu}G^{\text{AdS}}_{\frac{d}{2}+i\nu}({\hat X}_{{\cal A}_+},{\hat Y}_{{\cal A}_+}), \label{Feynasbubuads}
\end{align}
and 
\begin{align}\nonumber
G^{\left(m\right)}_{\Delta}(X_{{\cal A}_+ },Q)&=c_\Delta^{\text{dS-AdS}} \,\mathcal{K}^{(m)}_{i\left(\frac{d}{2}-\Delta\right)}( + e^{\frac{\pi i}{2}} R\,)\, i^ {+\Delta}K^{\text{AdS}}_{\Delta}\left(  s_{\text{AdS}}({\hat X}_{{\cal A}_+},Q)-i\epsilon\right),\\
G^{\left(m\right)}_{\Delta}(X_{{\cal D}_+},Q)&\to c^{\text{dS-AdS}}_{\Delta}\,\mathcal{K}^{(m)}_{i\left(\frac{d}{2}-\Delta\right)}(R\,)i^{- \Delta}\,K^{\text{AdS}}_{\Delta}(s_{\text{AdS}}({\hat X}_{{\cal A}_+},Q)-i\epsilon).\label{bubotoads}
\end{align}
Under the continuation \eqref{dsadscont} the integrals over region ${\cal D}_+$ can furthermore be re-cast as integrals over ${\cal A}_+$:
\begin{equation}
    \int_{{\cal D}_+} {\rm d}^{d+2}X \, \to \, e^{+ \frac{d \pi i}{2}}\int_{{\cal A}_+} {\rm d}^{d+2}X,
\end{equation}
where
\begin{align}
    \int_{{\cal D}_+} {\rm d}^{d+2}X \, &= \, \int^{\infty}_0 {\rm d}R\,R^{d+1}\,\int_{{\cal D}_+} {\rm d}^{d+1}{\hat X},\\
    \int_{{\cal A}_+}{\rm d}^{d+2}X &= \int^{\infty}_0 {\rm d}R\,R^{d+1}\,\int_{{\cal A}_+} {\rm d}^{d+1}{\hat X}.
\end{align}
All together the implication is that celestial correlation functions \eqref{ccdefn} can be perturbatively re-written in terms of corresponding Witten diagrams in EAdS \cite{Sleight:2023ojm,Iacobacci:2024nhw}, essentially following from the fact that the same is true \cite{Sleight:2020obc,Sleight:2021plv} for boundary correlators in dS.

\section{Conformal partial wave expansion}
\label{sec::CPWE}

The fact that celestial correlators \eqref{ccdefn} can be perturbatively recast in terms of corresponding Witten diagrams in EAdS suggests that we can in principle import techniques and results from the relatively well understood AdS case. In this work we consider the Conformal Partial Wave expansion (CPWE), which is the expansion of single-valued conformal correlation functions over an orthogonal basis of single-valued Eigenfunctions ${\cal F}_{\nu,J}$ of the Casimir invariants of the conformal group \cite{Dobrev:1975ru,Dobrev:1977qv,Mack:2009mi,Costa:2012cb,Caron-Huot:2017vep}. The CPWE of a conformal four-point function of operators ${\cal O}_{\Delta_i}$ with scaling dimensions $\Delta_i$ takes the form
\begin{equation}\label{cpwe}
    \langle {\cal O}_{\Delta_1}\left(Q_1\right){\cal O}_{\Delta_2}\left(Q_2\right){\cal O}_{\Delta_3}\left(Q_3\right){\cal O}_{\Delta_4}\left(Q_4\right) \rangle = \sum\limits_{J}\int^{\infty}_{-\infty}\frac{{\rm d}\nu}{2\pi}\,\rho_{J}\left(\nu\right){\cal F}_{\nu,J}\left(Q_1,Q_2,Q_3,Q_4\right),
\end{equation}
with spectral density $\rho_{J}\left(\nu\right)$. For boundary correlators in AdS, the spectral density $\rho_{J}\left(\nu\right)$ is a meromorphic function of $\nu$ which implies that they also admit a convergent expansion in terms of a discrete sum of conformal blocks ${\cal G}_{\Delta,J}$.\footnote{Conformal Blocks ${\cal G}_{\Delta,J}$ are Eigenfunctions of the Casimir invariants of the conformal group representing the contribution of an operator of arbitrary scale dimension $\Delta$ and spin $J$, together with its descendants, to a conformal four point function \cite{Dolan:2000ut,Dolan:2003hv,Dolan:2011dv}. Conformal Blocks are not single-valued functions and Conformal Partial Waves ${\cal F}_{\nu,J}$ are Eigenfunctions given by the single-valued linear combination \eqref{cpwascb} of Conformal Blocks.} This can be derived from the conformal partial wave expansion \eqref{cpwe} using that conformal partial waves are a specific (single-valued) linear combination of conformal blocks:\footnote{In this work we will only need these coefficients for spin-$J=0$, which are given explicitly by
\begin{equation}
    \kappa^{ij}_{\frac{d}{2}+i\nu,0} = \pi^{\frac{d}{2}}\frac{\Gamma\left(i\nu\right)}{\Gamma\left(\frac{d}{2}-i\nu\right)}\frac{\Gamma\left(\frac{\frac{d}{2}-i\nu+\Delta_i-\Delta_j}{2}\right)\Gamma\left(\frac{\frac{d}{2}-i\nu+\Delta_j-\Delta_i}{2}\right)}{\Gamma\left(\frac{\frac{d}{2}+i\nu+\Delta_i-\Delta_j}{2}\right)\Gamma\left(\frac{\frac{d}{2}+i\nu+\Delta_j-\Delta_i}{2}\right)}.
\end{equation}.}
\begin{multline}\label{cpwascb}
   {\cal F}^{12,34}_{\nu,J}(Q_1,Q_2,Q_3,Q_4) = \kappa^{34}_{\frac{d}{2}-i\nu,J}\, {\cal G}^{12,34}_{\frac{d}{2}+i\nu,J}(Q_1,Q_2,Q_3,Q_4)\\+\kappa^{12}_{\frac{d}{2}+i\nu,J} \, {\cal G}^{12,34}_{\frac{d}{2}-i\nu,J}(Q_1,Q_2,Q_3,Q_4),
\end{multline}
where the superscript $12,34$ refers to the expansion channel, in this case the $s$-channel. The conformal blocks ${\cal G}_{\frac{d}{2}\pm i\nu,J}$ decay exponentially as Im$\left(\nu\right)\to \mp \infty$ and by deforming the integration contour accordingly for each term one obtains the discrete expansion
\begin{equation}\label{cbe}
    \langle {\cal O}_{\Delta_1}\left(Q_1\right){\cal O}_{\Delta_2}\left(Q_2\right){\cal O}_{\Delta_3}\left(Q_3\right){\cal O}_{\Delta_4}\left(Q_4\right) \rangle = \sum_{\Delta^\prime}C_{\Delta^\prime,J}\,{\cal G}_{\Delta^\prime,J}\left(Q_1,Q_2,Q_3,Q_4\right),
\end{equation}
which is generated by the residues of the poles of $\rho_{J}\left(\nu\right)$ enclosed by the integration contour. The fact that celestial correlators can be rewritten in terms of EAdS Witten diagrams \cite{Sleight:2023ojm,Iacobacci:2024nhw} suggests that (at least perturbatively) they also admit a conformal partial wave expansion \eqref{cpwe} with meromorphic spectral density and, therefore, a conformal block expansion \eqref{cbe} as well. In this work we will show that this is indeed the case.

\vskip 4pt
A useful tool to determine the conformal partial wave expansion of Witten diagrams in AdS space is based on the spectral decomposition of bulk-to-bulk propagators in terms of an orthogonal basis of Eigenfunctions $\Omega_{\nu,J}$ of the Laplacian (see e.g. \cite{Penedones:2007ns,Costa:2014kfa}). The AdS harmonic function decomposition of the bulk-to-bulk propagator for a scalar field in EAdS$_{d+1}$ subject to the Dirichlet boundary condition i.e. $\Delta > \frac{d}{2}$ is:
\begin{align}\label{adsbubuharm}
    G^{\text{AdS}}_{\Delta}({\hat X},{\hat Y})&=\int^{\infty}_{-\infty}{\rm d}\nu\,a_{\Delta}\left(\nu\right)\Omega_{\nu,0}({\hat X},{\hat Y}), 
\end{align}
with spectral function 
\begin{equation}\label{sfbubuads}
    a_{\Delta}\left(\nu\right) = \frac{1}{\nu^2+\left(\Delta-\frac{d}{2}\right)^2}.
\end{equation}

For example, consider the four-point exchange Witten diagram in EAdS$_{d+1}$ generated by the cubic vertices
\begin{equation}\label{v3}
    {\cal V}_{12 \phi} = g_{12}\, \phi_1 \phi_2 \phi, \qquad {\cal V}_{34 \phi} = g_{34}\, \phi_3 \phi_4 \phi,
\end{equation}
with external scalar fields $\phi_i$ of mass $m_i^2=\Delta_i\left(\Delta_i-d\right)$ and exchanged field $\phi$ of mass $m^2=\Delta\left(\Delta-d\right)$ subject to the Dirichlet boundary condition. Applying the usual Feynman rules this reads
\begin{multline}
\label{AdSexch}\hspace*{-0.4cm}  {}^{{\cal V}_{12 \phi}{\cal V}_{34 \phi}}{\cal A}^{\text{AdS}}_{\Delta_1,\Delta_2|\Delta|\Delta_3,\Delta_4}\left(Q_1,Q_2,Q_3,Q_4\right) = g_{12}g_{34} \int_{{\cal A}_+} {\rm d}^{d+1}{\hat X}\,\int_{{\cal A}_+} {\rm d}^{d+1}{\hat Y}\,K^{\text{AdS}}_{\Delta_1}({\hat X},Q_1)K^{\text{AdS}}_{\Delta_2}({\hat X},Q_2)\\ \times G^{\text{AdS}}_{\Delta}({\hat X},{\hat Y})K^{\text{AdS}}_{\Delta_3}({\hat Y},Q_3)K^{\text{AdS}}_{\Delta_4}({\hat Y},Q_4).
\end{multline}
To determine the conformal partial wave expansion \eqref{cpwe} from the spectral decomposition \eqref{adsbubuharm} of the bulk-to-bulk propagator one notes that AdS Harmonic functions admit a so-called ``split representation", which expresses them as a product of AdS bulk-to-boundary propagators integrated over their common boundary point \cite{Leonhardt:2003qu,Moschella:2007zza,Costa:2014kfa}:
\begin{equation}\label{splitharm}
    \Omega_{\nu,0}({\hat X},{\hat Y}) = \frac{\nu^2}{\pi}\int_{\partial}{\rm d}Q\, K^{\text{AdS}}_{\frac{d}{2}+i\nu}({\hat X},Q)K^{\text{AdS}}_{\frac{d}{2}-i\nu}({\hat Y},Q).
\end{equation}
This reduces the AdS integrals to three-point contact Witten diagrams \cite{Freedman:1998tz}:
\begin{align}
  {}^{{\cal V}_{123}}{\cal A}^{\text{AdS}}_{\Delta_1,\Delta_2,\Delta_3}\left(Q_1,Q_2,Q_3\right) &= \int_{{\cal A}_+} {\rm d}^{d+1}{\hat X}\,K^{\text{AdS}}_{\Delta_1}({\hat X},Q_1)K^{\text{AdS}}_{\Delta_2}({\hat X},Q_2)K^{\text{AdS}}_{\Delta_3}({\hat X},Q_3),\\
  &= C^{\text{AdS}}_{\Delta_1\Delta_2\Delta_3} \langle \langle {\cal O}_{\Delta_1}\left(Q_1\right){\cal O}_{\Delta_2}\left(Q_2\right){\cal O}_{\Delta_3}\left(Q_3\right) \rangle \rangle,
\end{align}
where the notation $\langle \langle \bullet \rangle \rangle$ denotes the unit normalised conformal three-point structure
\begin{multline}
  \langle \langle {\cal O}_{\Delta_1}\left(Q_1\right){\cal O}_{\Delta_2}\left(Q_2\right){\cal O}_{\Delta_3}\left(Q_3\right) \rangle \rangle \\ = \frac{1}{\left(-2 Q_1 \cdot Q_2\right)^{\frac{\Delta_1+\Delta_2-\Delta_3}{2}}\left(-2 Q_1 \cdot Q_3\right)^{\frac{\Delta_1+\Delta_3-\Delta_2}{2}}\left(-2 Q_2 \cdot Q_3\right)^{\frac{\Delta_2+\Delta_3-\Delta_1}{2}}},
\end{multline}
and the three-point coefficient is
\begin{multline}
    C^{\text{AdS}}_{\Delta_1 \Delta_2 \Delta_3} = \frac{1}{2} \pi^{\frac{d}{2}}\, \frac{C^{\text{AdS}}_{\Delta_1}C^{\text{AdS}}_{\Delta_2}C^{\text{AdS}}_{\Delta_3}}{\Gamma\left(\Delta_1\right)\Gamma\left(\Delta_2\right)\Gamma\left(\Delta_3\right)} \Gamma\left(\frac{\Delta_1+\Delta_2+\Delta_3-d}{2}\right)\\ \times \Gamma\left(\frac{\Delta_1+\Delta_2-\Delta_3}{2}\right)\Gamma\left(\frac{\Delta_1+\Delta_3-\Delta_2}{2}\right)\Gamma\left(\frac{\Delta_2+\Delta_3-\Delta_1}{2}\right).
\end{multline}
The utility of the split representation \eqref{splitharm} stems from the fact that conformal partial waves admit an analogous integral representation \cite{Dolan:2011dv}:
\begin{multline}
    {\cal F}^{12,34}_{\nu,0}(Q_1,Q_2,Q_3,Q_4) \\ = \int_{\partial}{\rm d}Q\, \langle \langle {\cal O}_{\Delta_1}\left(Q_1\right){\cal O}_{\Delta_2}\left(Q_2\right){\cal O}_{\frac{d}{2}+i\nu}\left(Q\right) \rangle \rangle \langle \langle {\cal O}_{\frac{d}{2}-i\nu}\left(Q\right) {\cal O}_{\Delta_3}\left(Q_3\right){\cal O}_{\Delta_4}\left(Q_4\right) \rangle \rangle,
\end{multline}
from which one can read off the conformal partial wave expansion for the exchange diagram \eqref{AdSexch} to be:
\begin{multline}
   {}^{{\cal V}_{12 \phi}{\cal V}_{34 \phi}}{\cal A}^{\text{AdS}}_{\Delta_1,\Delta_2|\Delta|\Delta_3,\Delta_4}\left(Q_1,Q_2,Q_3,Q_4\right) = \int^{\infty}_{-\infty}\frac{{\rm d}\nu}{2\pi}\,\rho_{\Delta_1,\Delta_2|\Delta|\Delta_3,\Delta_4}\left(\nu\right)\\ \times {\cal F}^{12,34}_{\nu,0}\left(Q_1,Q_2,Q_3,Q_4\right),
\end{multline}
with spectral density 
\begin{align}\label{adsexchsd}
    \rho_{\Delta_1,\Delta_2|\Delta|\Delta_3,\Delta_4}\left(\nu\right) &= a_{\Delta}\left(\nu\right) \, \rho^{\text{AdS}}_{\Delta_1,\Delta_2\Delta_3,\Delta_4}\left(\nu\right).
\end{align}
The factor $a_{\Delta}\left(\nu\right)$ originates from the propagator \eqref{adsbubuharm} for the exchanged field while the factor
\begin{equation}
    \rho^{\text{AdS}}_{\Delta_1,\Delta_2\Delta_3,\Delta_4}\left(\nu\right)=2\nu^2C^{\text{AdS}}_{\Delta_1 \Delta_2 \frac{d}{2}+i\nu}C^{\text{AdS}}_{\frac{d}{2}-i\nu \Delta_3 \Delta_4}, \label{adsexchrho}
\end{equation}
encodes the contribution from the integration over the bulk of AdS.

\vskip 4pt
The spectral density \eqref{adsexchsd} is a meromorphic function of $\nu$, which implies that the exchange can be expanded in a discrete sum \eqref{cbe} of conformal blocks. In this case it also has the symmetry property\footnote{In the case that this property does not hold, extra care must be taken with the integration contour to bring the conformal partial wave expansion into the form \eqref{exchintcb} as an integral of a single conformal block. We will encounter an example of this type in section \ref{subsec::exchdiagramecbw}.}
\begin{equation}\label{sdsymm}
     \rho_{\Delta_1,\Delta_2|\Delta|\Delta_3,\Delta_4}\left(-\nu\right)\kappa^{12}_{\frac{d}{2}-i\nu,0} =  \rho_{\Delta_1,\Delta_2|\Delta|\Delta_3,\Delta_4}\left(\nu\right)\kappa^{34}_{\frac{d}{2}-i\nu,0},
\end{equation}
so that, inserting the expression \eqref{cpwascb} for a conformal partial wave as a sum of two conformal blocks, we can write:
\begin{multline}\label{exchintcb}
   {}^{{\cal V}_{12 \phi}{\cal V}_{34 \phi}}{\cal A}^{\text{AdS}}_{\Delta_1,\Delta_2|\Delta|\Delta_3,\Delta_4}\left(Q_1,Q_2,Q_3,Q_4\right)\\ = \int^{\infty}_{-\infty}\frac{{\rm d}\nu}{2\pi}\,\rho_{\Delta_1,\Delta_2|\Delta|\Delta_3,\Delta_4}\left(\nu\right)2\kappa^{34}_{\frac{d}{2}-i\nu,0}\,{\cal G}^{12,34}_{\frac{d}{2}+i\nu,0}\left(Q_1,Q_2,Q_3,Q_4\right).
\end{multline}
Since ${\cal G}^{12,34}_{\frac{d}{2}+i\nu,0}$ decays exponentially as $\text{Im}\left(\nu\right) \to -\infty$, the conformal block expansion is generated by pulling the contour downwards and evaluating the residues of poles in the lower half plane. From the spectral function \eqref{sfbubuads} for the bulk-to-bulk propagator we have at pole at
\begin{equation}\label{STpoleAdS}
    \frac{d}{2}+i\nu = \Delta,
\end{equation}
which encodes the exchanged single-particle state in AdS with scaling dimension $\Delta$. The bulk integrals over AdS give the spectral function \eqref{adsexchrho} and in the lower-half plane this contributes two infinite families of poles:
\begin{subequations}\label{DTAdS}
  \begin{align}
    \frac{d}{2}+i\nu&=\Delta_1+\Delta_2+2n, \qquad n =0, 1, 2, \ldots, \\
    \frac{d}{2}+i\nu&=\Delta_3+\Delta_4+2n, \qquad n =0, 1, 2, \ldots.
\end{align}  
\end{subequations}
These correspond to composite (``double-trace") operators which have the schematic form:
\begin{equation}\label{DTops}
    \left[{\cal O}_{\Delta_1}{\cal O}_{\Delta_2}\right]_{n}\sim {\cal O}_{\Delta_1} (\partial^ 2)^ n {\cal O}_{\Delta_2}+\ldots, \qquad \left[{\cal O}_{\Delta_3}{\cal O}_{\Delta_4}\right]_{n}\sim {\cal O}_{\Delta_3} (\partial^ 2)^ n {\cal O}_{\Delta_4}+\ldots\,.
\end{equation}

In the following we extend the AdS harmonic function decomposition to the Feynman propagator in Minkowski space. This will allow to extend the above tools to celestial correlators \eqref{ccdefn}.

\subsection{AdS harmonic function decomposition of the Feynman propagator}
\label{subsec::adsfeyn}

We can now extend the AdS harmonic function decomposition to the Feynman propagator in $\mathbb{M}^{d+2}$ using the decomposition \eqref{Feynasbubuads} of the latter in terms of bulk-to-bulk propagators in EAdS$_{d+1}$. We first note that, using the symmetries of the spectral integral in this decomposition, the Feynman propagator can be expressed in the form (see appendix \ref{app::FeynAdS}): 
\begin{align} \nonumber
\hspace*{-0.75cm}G^{(m)}_T(X_{{\cal A}_+},Y_{{\cal A}_+})&=\int_{-\infty}^{+\infty}\frac{{\rm d}\nu}{2\pi}\,2i\nu\, g_{-\tfrac{d}{2}+i\nu}^{(m)}(e^{\frac{\pi i}{2}}R_1,e^{\frac{\pi i}{2}}R_2) \left[ e^{\left(\frac{d}{2}+i\nu\right)\pi i}c^{\text{dS-AdS}}_{\frac{d}{2}+i\nu} G^{\text{AdS}}_{\frac{d}{2}+i\nu}({\hat X}_{{\cal A}_+},{\hat Y}_{{\cal A}_+})+\left(\nu \to -\nu\right)\right],\\ \nonumber
 \hspace*{-0.75cm}  G^{(m)}_T(X_{{\cal D}_+},Y_{{\cal D}_+})
   &\to \int_{-\infty}^{+\infty}\frac{{\rm d}\nu}{2\pi}\,2i\nu\,g_{-\tfrac{d}{2}+i\nu}^{(m)}(R_1,R_2)\left[e^{-(\frac{d}{2}+i\nu)\pi i}c^{\text{dS-AdS}}_{\frac{d}{2}+i\nu}G^{\text{AdS}}_{\frac{d}{2}+i\nu}({\hat X}_{{\cal A}_+},{\hat Y}_{{\cal A}_+})+\left(\nu \to -\nu\right)\right],\\ \label{Feynsymads}
 \hspace*{-01cm}  G^{(m)}_T(X_{{\cal D}_+},Y_{{\cal A}_+})&\to \int_{-\infty}^{+\infty}\frac{{\rm d}\nu}{2\pi}\,2i\nu\,g_{-\tfrac{d}{2}+i\nu}^{(m)}(R_1,e^ {\frac{\pi i}{2}}R_2)\left[c^{\text{dS-AdS}}_{\frac{d}{2}+i\nu}G^{\text{AdS}}_{\frac{d}{2}+i\nu}({\hat X}_{{\cal A}_+},{\hat Y}_{{\cal A}_+})+\left(\nu \to -\nu\right)\right],
\end{align}
where the spectral function \eqref{rho} has been replaced by 
\begin{multline}\label{bubug}
    g_{-\tfrac{d}{2}+i\nu}^{(m)}(R_1,R_2)= R^{-\frac{d}{2}}_1R^{-\frac{d}{2}}_2\Big[\theta(R_2-R_1) K_{i\nu}\left(m R_1\right)I_{i\nu}\left(m R_2\right)\\+\theta(R_1-R_2) K_{i\nu}\left(m R_2\right)I_{i\nu}\left(m R_1\right)\Big],
\end{multline}
where $I_{i\nu}$ and $K_{i\nu}$ are modified Bessel functions of the first and second kind. Interestingly, like the radial dependence \eqref{rkernel} of the celestial bulk-to-boundary propagator, the radial function \eqref{bubug} takes the form of the momentum space bulk-to-bulk propagator in EAdS$_{1-d}$ for a scalar field with scaling dimension $-\tfrac{d}{2}+i\nu$ \cite{Liu:1998ty}, where the mass $m$ plays the role of the modulus of the boundary momentum and $R_1, R_2$ the bulk Poincar\'e coordinates.\footnote{A similar observation was made in \cite{Iacobacci:2024nhw}, section 5.}  

\vskip 4pt 
The key observation is that the bulk-to-bulk propagator $G^{\text{AdS}}_{\frac{d}{2}+i\nu}$ is holomorphic in the lower-half $\nu$-plane, so that terms proportional it in the square bracket of the spectral integral \eqref{Feynsymads} give a vanishing contribution. Combining this observation with the well known identity \cite{Costa:2014kfa}:
\begin{equation}
G^{\text{AdS}}_{\frac{d}{2}+i\nu}({\hat X}_{{\cal A}_+},{\hat Y}_{{\cal A}_+})-G^{\text{AdS}}_{\frac{d}{2}-i\nu}({\hat X}_{{\cal A}_+},{\hat Y}_{{\cal A}_+})=\frac{2\pi}{i\nu}\Omega_{\nu,0}({\hat X}_{{\cal A}_+},{\hat Y}_{{\cal A}_+}),
\end{equation}
yields the following decomposition of the Minkowski Feynman propagator in terms of AdS harmonic functions in the different regions of the hyperbolic slicing:
\begin{shaded} 
\noindent \emph{AdS harmonic function decomposition of the Feynman propagator.}
 \begin{align} \nonumber
\hspace*{-0.75cm}G^{(m)}_T(X_{{\cal A}_+},Y_{{\cal A}_+})&=4\pi \int_{-\infty}^{+\infty}\frac{{\rm d}\nu}{2\pi}\, g_{-\tfrac{d}{2}+i\nu}^{(m)}(e^{\frac{\pi i}{2}}R_1,e^{\frac{\pi i}{2}}R_2)  e^{\left(\frac{d}{2}-i\nu\right)\pi i}c^{\text{dS-AdS}}_{\frac{d}{2}+i\nu} \Omega_{\nu,0}({\hat X}_{{\cal A}_+},{\hat Y}_{{\cal A}_+}),\\ \nonumber
 \hspace*{-0.75cm}  G^{(m)}_T(X_{{\cal D}_+},Y_{{\cal D}_+})
   &\to 4\pi \int_{-\infty}^{+\infty}\frac{{\rm d}\nu}{2\pi}\,g_{-\tfrac{d}{2}+i\nu}^{(m)}(R_1,R_2)e^{-(\frac{d}{2}-i\nu)\pi i}c^{\text{dS-AdS}}_{\frac{d}{2}+i\nu}\Omega_{\nu,0}({\hat X}_{{\cal A}_+},{\hat Y}_{{\cal A}_+}),\\ 
 \hspace*{-0.75cm}  G^{(m)}_T(X_{{\cal D}_+},Y_{{\cal A}_+})&\to 4\pi \int_{-\infty}^{+\infty}\frac{{\rm d}\nu}{2\pi}\,g_{-\tfrac{d}{2}+i\nu}^{(m)}(R_1,e^ {\frac{\pi i}{2}}R_2)c^{\text{dS-AdS}}_{\frac{d}{2}+i\nu}\Omega_{\nu,0}({\hat X}_{{\cal A}_+},{\hat Y}_{{\cal A}_+}). \label{bubuharm}
\end{align}   
\end{shaded}
These can now be applied to determine the conformal partial wave expansion \eqref{cpwe} of perturbative celestial correlators \eqref{ccdefn}, similar to the AdS case reviewed above. The spectral density is obtained upon evaluating the integrals over the radial direction of Minkowski space which, interestingly, take the form of the corresponding momentum space Witten diagram in EAdS$_{1-d}$. This follows from the structure \eqref{bubug} of the spectral functions for the Feynman propagator and that of the radial kernel \eqref{rkernel} for the celestial bulk-to-boundary propagator. In the next section this is discussed in more detail for tree level exchange diagrams in $\mathbb{M}^{d+2}$.

\subsection{Example: Tree level exchange diagrams.} 
\label{subsecc::TLexch}

In this section we apply the decomposition of the Minkowski Feyman propagator in terms of AdS harmonic functions to determine the conformal partial wave expansion of tree-level exchange diagram contributions to celestial correlators \eqref{ccdefn}.

\vskip 4pt
Consider the four-point exchange diagram in $\mathbb{M}^{d+2}$ generated by the cubic vertices
\begin{equation}
    {\cal V}_{12 \phi} = g_{12}\, \phi_1 \phi_2 \phi, \qquad {\cal V}_{34 \phi} = g_{34}\, \phi_3 \phi_4 \phi,
\end{equation}
with external scalar fields $\phi_i$ of mass $m_i$ and the exchanged field $\phi$ has mass $m$. Applying the Feynman rules, this is given by
\begin{multline}
  \hspace*{-0.5cm} {\cal A}^{(m_1,m_2|m|m_3,m_4)}_{\Delta_1 \Delta_2\Delta_3 \Delta_4}\left(Q_1,\ldots,Q_4\right)=\left(-ig_{12}\right)\left(-ig_{34}\right) \int {\rm d}^{d+2}X\,\int {\rm d}^{d+2}Y\,G^{(m_1)}_{\Delta_1}\left(X,Q_1\right)G^{(m_2)}_{\Delta_2}\left(X,Q_2\right)\\ \times G^{\left(m\right)}_{T}\left(X,Y\right)G^{(m_3)}_{\Delta_3}\left(Y,Q_3\right)G^{(m_4)}_{\Delta_4}\left(Y,Q_4\right). \label{celexch}
\end{multline}

From the harmonic function decomposition \eqref{bubuharm} of the Feynman propagator and the expressions \eqref{bubotoads} for the celestial bulk-to-boundary propagators in terms of those in EAdS, we can write down the $s$-channel conformal partial wave expansion as (see appendix \ref{app::cpweexch}) 
\begin{multline}
   {\cal A}^{(m_1,m_2|m|m_3,m_4)}_{\Delta_1 \Delta_2 \Delta_3 \Delta_4}\left(Q_1,Q_2,Q_3,Q_4\right)\\ = \int^{\infty}_{-\infty}\frac{{\rm d}\nu}{2\pi}\,\rho^{(m_1,m_2|m|m_3,m_4)}_{\Delta_1 \Delta_2 \Delta_3 \Delta_4}\left(\nu\right){\cal F}^{12,34}_{\nu,0}\left(Q_1,Q_2,Q_3,Q_4\right),
\end{multline}
with spectral density 
\begin{equation}\label{spectcexch}
  \hspace*{-1cm}  \rho^{(m_1,m_2|m|m_3,m_4)}_{\Delta_1 \Delta_2 \Delta_3 \Delta_4}\left(\nu\right)= {\bar \rho}^{(m_1,m_2|m|m_3,m_4)}_{\Delta_1 \Delta_2 \Delta_3 \Delta_4}\left(\nu\right)
  \rho^{\text{AdS}}_{\Delta_1 \Delta_2 \Delta_3 \Delta_4}\left(\nu\right),
\end{equation}
where it is useful to separate out the contribution \eqref{adsexchrho} from the co-dimension one EAdS hyperboloid,  defining a reduced spectral density:
\begin{multline}\label{spectcexchbar}
    {\bar \rho}^{(m_1,m_2|m|m_3,m_4)}_{\Delta_1 \Delta_2 \Delta_3 \Delta_4}\left(\nu\right)=16 \pi \left(\prod^4_{i=1}c^{\text{dS-AdS}}_{\Delta_i} \right)\sin\left(\tfrac{\pi}{2}\left(-\tfrac{d}{2}-i\nu+\Delta_1+\Delta_2\right)\right)\sin\left(\tfrac{\pi}{2}\left(-\tfrac{d}{2}-i\nu+\Delta_3+\Delta_4\right)\right)\\ \times c^{\text{dS-AdS}}_{\frac{d}{2}+i\nu}\, a^{(m_1,m_2|m|m_3,m_4)}_{\Delta_1,\Delta_2|-\tfrac{d}{2}+i\nu|\Delta_3,\Delta_4}, 
\end{multline}
where
\begin{multline}\label{aexch}
  a^{(m_1,m_2|m|m_3,m_4)}_{\Delta_1,\Delta_2|-\tfrac{d}{2}+i\nu|\Delta_3,\Delta_4} = \int^\infty_0 {\rm d}R_1 {\rm d}R_2 R^{d+1}_1 R^{d+1}_2\,\mathcal{K}^{(m_1)}_{i\left(\frac{d}{2}-\Delta_1\right)}(R_1)\mathcal{K}^{(m_2)}_{i\left(\frac{d}{2}-\Delta_2\right)}(R_1)\\ \times g^{(m)}_{-\tfrac{d}{2}+i\nu}(R_1,R_2)\mathcal{K}^{(m_3)}_{i\left(\frac{d}{2}-\Delta_3\right)}(R_2)\mathcal{K}^{(m_4)}_{i\left(\frac{d}{2}-\Delta_4\right)}(R_2),
\end{multline}
which encodes the contribution from the radial direction in the hyperbolic slicing of Minkowski space. As can be understood in more detail in appendix \ref{app::cpweexch}, the sinusoidal factors in \eqref{spectcexchbar} arise from combining the contributions from each region ${\cal A}_+$ and ${\cal D}_+$ of the hyperbolic slicing -- which differ by a phase. 

\vskip 4pt
Unlike for the exchange Witten diagram \eqref{cpwecexch} in AdS, the spectral function \eqref{spectcexch} does not satisfy the symmetry property \eqref{sdsymm} owing to the asymmetry of the reduced spectral density \eqref{spectcexchbar} in $\nu$. It is still however a meromorphic function of $\nu$ and therefore admits a conformal block expansion. Following from the radial decomposition \eqref{spectralrepcelesybubu}, the integral over $\nu$ in \eqref{cpwecexch} should be regarded as a Mellin-Barnes type integral where the integration contour is usually a deformation of the real axis such that poles encoded in $\Gamma$-function factors of the form $\Gamma\left(a+i\nu\right)$ are separated from those of the form $\Gamma\left(b-i\nu\right)$. The conformal block expansion is then obtained by inserting the expression \eqref{cpwascb} for the conformal partial wave as a sum of two conformal blocks and pulling the integration contour down for the term involving ${\cal G}^{12,34}_{\frac{d}{2}+i\nu,0}$ and up for the term involving ${\cal G}^{12,34}_{\frac{d}{2}-i\nu,0}$. 

\vskip 4pt
On the other hand, if due care is taken with the integration contour, it is possible to obtain an expression for the conformal partial wave expansion in a more familiar form with a spectral function satisfying the symmetry property \eqref{sdsymm}. In particular, assuming that no poles of the reduced spectral function \eqref{spectcexchbar} cross the real axis one can write:
\begin{multline}\label{cpwecexch}
   {\cal A}^{(m_1,m_2|m|m_3,m_4)}_{\Delta_1 \Delta_2 \Delta_3 \Delta_4}\left(Q_1,Q_2,Q_3,Q_4\right) = \int^{\infty}_{-\infty}\frac{{\rm d}\nu}{2\pi}\,\left[{\bar \rho}^{(m_1,m_2|m|m_3,m_4)}_{\Delta_1 \Delta_2 \Delta_3 \Delta_4}\left(\nu\right)+{\bar \rho}^{(m_1,m_2|m|m_3,m_4)}_{\Delta_1 \Delta_2 \Delta_3 \Delta_4}\left(-\nu\right)\right]\\ \times  \rho^{\text{AdS}}_{\Delta_1 \Delta_2 \Delta_3 \Delta_4}\left(\nu\right){\cal F}^{12,34}_{\nu,0}\left(Q_1,Q_2,Q_3,Q_4\right),
\end{multline}
where the spectral function multiplying the conformal partial wave now satisfies the property \eqref{sdsymm}. In the case that there are poles in \eqref{spectcexchbar} crossing the real axis, the two terms in the square bracket would not share the same integration contour. Obtaining a symmetric spectral function in this case would require subtractions to account for the different poles captured by the different integration contours. When referring to this form for the conformal partial wave expansion in the general discussion we shall keep any required subtractions implicit. 

\vskip 4pt
Before discussing the corresponding expansion into conformal blocks let us note that, as anticipated at the end of section \ref{subsec::adsfeyn}, the function \eqref{aexch} encoding the radial contributions takes the form of the momentum space exchange Witten diagram in EAdS$_{1-d}$ for a scalar field of scaling dimension $-\frac{d}{2}+i\nu$. The radial kernels \eqref{rkernel} play the role of bulk-to-boundary propagators, the function \eqref{bubug} the role of the bulk-to-bulk propagator and the masses $m_i$ and $m$ the role of the modulus of the boundary momenta. With the latter identifications it is clear that this observation holds more generally: For any given perturbative contribution to celestial correlators \eqref{ccdefn}, at the level of the harmonic function decomposition \eqref{bubuharm} for the Feynman propagator the radial integrals take the form of the corresponding momentum space Witten diagram in EAdS$_{1-d}$.

\paragraph{Conformal block expansion.} Since the spectral function \eqref{spectcexch} is a meromorphic function of $\nu$, the celestial exchange \eqref{celexch} also admits an expansion \eqref{cbe} into conformal blocks. Inserting the expression \eqref{cpwascb} for the conformal partial wave in terms of conformal blocks we have: 
\begin{multline}\label{Fexch2cb}
   {\cal A}^{(m_1,m_2|m|m_3,m_4)}_{\Delta_1 \Delta_2 \Delta_3 \Delta_4}\left(Q_1,Q_2,Q_3,Q_4\right) = \int^{\infty}_{-\infty}\frac{{\rm d}\nu}{2\pi}\,\rho^{(m_1,m_2|m|m_3,m_4)}_{\Delta_1 \Delta_2 \Delta_3 \Delta_4}\left(\nu\right)\\ \times  \left[\kappa^{34}_{\frac{d}{2}-i\nu,0}{\cal G}^{12,34}_{\frac{d}{2}+i\nu,0}\left(Q_1,Q_2,Q_3,Q_4\right)+\kappa^{12}_{\frac{d}{2}+i\nu,0}{\cal G}^{12,34}_{\frac{d}{2}-i\nu,0}\left(Q_1,Q_2,Q_3,Q_4\right)\right].
\end{multline}
The conformal blocks ${\cal G}^{12,34}_{\frac{d}{2}\pm i\nu,0}$ decay exponentially as Im$\left(\nu\right)\to \mp \infty$ and so the conformal block expansion is generated by pulling the integration contour down for the term involving ${\cal G}^{12,34}_{\frac{d}{2}+i\nu,0}$ and up for the term involving ${\cal G}^{12,34}_{\frac{d}{2}-i\nu,0}$. Alternatively, assuming that there are no poles in the reduced spectral function \eqref{rhobar} that cross the real axis, one can start from the more familiar symmetric form \eqref{cpwecexch} for the conformal partial wave expansion, giving: 
\begin{multline}\label{Fexch2cbsymm}
  \hspace*{-0.4cm} {\cal A}^{(m_1,m_2|m|m_3,m_4)}_{\Delta_1 \Delta_2 \Delta_3 \Delta_4}\left(Q_1,Q_2,Q_3,Q_4\right) = \int^{\infty}_{-\infty}\frac{{\rm d}\nu}{2\pi}\,\left[{\bar \rho}^{(m_1,m_2|m|m_3,m_4)}_{\Delta_1 \Delta_2 \Delta_3 \Delta_4}\left(\nu\right)+{\bar \rho}^{(m_1,m_2|m|m_3,m_4)}_{\Delta_1 \Delta_2 \Delta_3 \Delta_4}\left(-\nu\right)\right]\\ \times \rho^{\text{AdS}}_{\Delta_1 \Delta_2 \Delta_3 \Delta_4}\left(\nu\right)2\kappa^{34}_{\frac{d}{2}-i\nu,0}{\cal G}^{12,34}_{\frac{d}{2}+i\nu,0}\left(Q_1,Q_2,Q_3,Q_4\right),
\end{multline}
where for convenience we repeat the definition of the reduced spectral density below:
\begin{multline}\label{spectcexchbar2}
 \hspace*{-1.25cm}   {\bar \rho}^{(m_1,m_2|m|m_3,m_4)}_{\Delta_1 \Delta_2 \Delta_3 \Delta_4}\left(\nu\right)=16 \pi \left(\prod^4_{i=1}c^{\text{dS-AdS}}_{\Delta_i} \right)\sin\left(\tfrac{\pi}{2}\left(-\tfrac{d}{2}-i\nu+\Delta_1+\Delta_2\right)\right)\sin\left(\tfrac{\pi}{2}\left(-\tfrac{d}{2}-i\nu+\Delta_3+\Delta_4\right)\right)\\ \times c^{\text{dS-AdS}}_{\frac{d}{2}+i\nu}\, a^{(m_1,m_2|m|m_3,m_4)}_{\Delta_1,\Delta_2|-\tfrac{d}{2}+i\nu|\Delta_3,\Delta_4}. 
\end{multline}

\vskip 4pt
One can already conclude that the conformal block expansion contains contributions from the two infinite families of double-trace operators \eqref{DTops}, like in AdS. These are encoded in the function $\rho^{\text{AdS}}_{\Delta_1 \Delta_2 \Delta_3 \Delta_4}\left(\nu\right)$, which arises from the integral over the co-dimension one EAdS hyperboloid. Interestingly, the sinusodial factors from the reduced spectral density \eqref{spectcexchbar2} cancel the double-trace poles in the lower-half plane. The double-trace contributions \eqref{DTops} are therefore only generated from the term in the conformal partial wave expansion \eqref{Fexch2cb} with conformal block ${\cal G}^{12,34}_{\frac{d}{2}-i\nu,0}$, for which one closes in the upper-half plane. In the symmetric form \eqref{Fexch2cbsymm} of the conformal partial wave expansion, this corresponds to the term with $\nu \to -\nu$ in the square brackets.

\vskip 4pt
The remaining contributions to the conformal block expansion are specific to celestial exchange diagrams and are encoded in the reduced spectral function \eqref{rhobar}. The non-trivial contributions\footnote{There are also poles from the the factor $c^{\text{dS-AdS}}_{\frac{d}{2}+i\nu}$ in \eqref{spectcexchbar2}, given explicitly by \eqref{dS-AdScoeff}. These however do not generate a non-trivial contribution to the conformal block expansion since, being an odd function of $\nu$, their residues from the term with ${\cal G}^{12,34}_{\frac{d}{2}+i\nu,0}$ cancel with those from ${\cal G}^{12,34}_{\frac{d}{2}-i\nu,0}$. At the level of the symmetric representation \eqref{Fexch2cbsymm}, the square bracket is vanishing on these residues.} are all generated by the function \eqref{aexch}, 
\begin{multline}
  a^{(m_1,m_2|m|m_3,m_4)}_{\Delta_1,\Delta_2|-\tfrac{d}{2}+i\nu|\Delta_3,\Delta_4} = \int^\infty_0 {\rm d}R_1 {\rm d}R_2 R^{d+1}_1 R^{d+1}_2\,\mathcal{K}^{(m_1)}_{i\left(\frac{d}{2}-\Delta_1\right)}(R_1)\mathcal{K}^{(m_2)}_{i\left(\frac{d}{2}-\Delta_2\right)}(R_1)\\ \times g^{(m)}_{-\tfrac{d}{2}+i\nu}(R_1,R_2)\mathcal{K}^{(m_3)}_{i\left(\frac{d}{2}-\Delta_3\right)}(R_2)\mathcal{K}^{(m_4)}_{i\left(\frac{d}{2}-\Delta_4\right)}(R_2),
\end{multline}
which arises from the integral over the radial direction in the hyperbolic slicing. For generic masses $m_i$ its explicit form is quite involved (see \eqref{agenm} in appendix \ref{app::cpweexch}), though there are significant simplifications for correlators of massless fields. This case is considered in more detail in the following sections.

\section{Massless celestial correlators}
\label{sec::MLcase}

In this section we consider celestial correlation functions \eqref{ccdefn} for massless external scalar fields. The celestial bulk-to-boundary propagator \eqref{cbuboexpl} for massless scalar fields can be obtained from the limit $m\to0$ of the massive case \eqref{cbuboexpl}: Writing $\Delta = \tfrac{d}{2}+i\left(\nu+i\epsilon\right)$ with $\epsilon >0$ and expanding the radial kernel \eqref{rkernel} for $m\to 0$, we have 
\begin{align}
    \mathcal{K}^{(m)}_{\nu}(R) &= R^{\Delta-d}\left[\left(1+O\left(m\right)\right)+\frac{\Gamma\left(\Delta-\tfrac{d}{2}\right)}{\Gamma\left(\tfrac{d}{2}-\Delta\right)}\left(\frac{m}{2}\right)^{-i\nu+2\epsilon}\left(1+O\left(m\right)\right)\right],\\
    &\overset{m\to0}{\to} R^{\Delta-d}, \label{bubopl}
\end{align}
which gives the celestial bulk-to-boundary propagator for massless scalar fields as
\begin{align}\label{cbuboml}
    G^{(0)}_{\Delta}(X,Q)&:=\lim_{m\to 0} G^{(m)}_{\Delta}(X,Q),\\
    &=G_\Delta(\hat{X}_\epsilon,Q)\left(\sqrt{X^2+i\epsilon}\right)^{\Delta-d}.\label{cbuboml2}
\end{align}
This expression can also be obtained directly from the Mellin transform \eqref{celestialbubo} of the massless Feynman propagator (see appendix \ref{A}), meaning that the massless limit commutes with the Mellin transform; the massless limit is smooth, as it is for the Feynman propagator. 

\vskip 4pt
It is interesting to compare the celestial bulk-to-boundary propagators \eqref{cbuboml} with conformal primary wave functions for celestial amplitudes \cite{Pasterski:2016qvg,Pasterski:2017kqt}. In particular, we note that the expression \eqref{cbuboml2} for massless scalars corresponds to the shadow conformal primary basis for massless scalar fields (see equation (4.10) of \cite{Pasterski:2017kqt}), modulo normalisation and $+i\epsilon$ prescription accounting for the different choice of observable (i.e. time ordered correlation functions as opposed to S-matrix elements). For a related discussion see \cite{Jorstad:2023ajr}. For celestial amplitudes \cite{Pasterski:2016qvg,Pasterski:2017kqt}, a smooth massless limit requires a shadow transformation of conformal primary wave functions for incoming particles \cite{Furugori:2023hgv}.

\vskip 4pt
In the following we explore in more detail some of the properties of celestial correlation functions for massless scalars. The fact that the radial dependence of massless celestial bulk-to-boundary propagators is a power law \eqref{bubopl} significantly simplifies the evaluation of radial integrals in this case since they correspond to Dirac delta functions in the scaling dimensions. This gives rise to some interesting features of massless celestial correlators compared to their massive counterparts. 

\vskip 4pt
Let us first note that some interesting properties already emerge at the level of the free theory celestial two-point function, where the massless limit of \eqref{bdry2pt} gives
\begin{align}
G_{\Delta_1\Delta_2}^{(0)}(Q_1,Q_2)&=\lim_{m\to0}G_{\Delta_1\Delta_2}^{(m)}(Q_1,Q_2),\\ \nonumber
&=\left(\frac{1}{2}\right)^{d-2\Delta_1}\frac{\Gamma(\Delta_1)}{4\pi^{\frac{d+2}{2}}(-2Q_1\cdot Q_2+i\epsilon)^{\Delta_1}}(2\pi)\delta(i(\Delta_1-\Delta_2))\\
&\hspace*{3cm}\times\underbrace{\lim_{m\to0}\left[m^{-2(\Delta_1-\frac{d}{2})}\Gamma\left(\Delta_1-\frac{d}{2}\right)\right]}_{=(2\pi i)\delta\left(\Delta_1-\frac{d}{2}\right)},\\
&=\frac{\Gamma(\Delta_1)}{4\pi^{\frac{d+2}{2}}}\frac{1}{(-2Q_1\cdot Q_2+i\epsilon)^{\Delta_1}}(2\pi i)\delta(\Delta_1-\Delta_2)(2\pi i)\delta\left(\Delta_1-\tfrac{d}{2}\right). \label{c2ptml}
\end{align}
This can be equivalently obtained from the double Mellin transform of the massless Feynman propagator (see appendix \ref{C}). The two-point function therefore only has support on $\Delta_1=\Delta_2=\frac{d}{2}$ in the massless case. A similar feature has been observed in the context of the Carrollian approach to flat space holography, where the free theory two-point function for massless scalars at null infinity also only has support for these values of the scaling dimension \cite{Nguyen:2023miw}. In both cases this feature likely originates from the power-law behaviour \eqref{appmlFeyn} of the massless Feynman propagator.

\subsection{Contact diagrams}

In this section we consider for completeness contact diagram contributions to celestial correlators \eqref{ccdefn} involving massless scalars. Celestial contact diagrams are proportional to contact Witten diagrams in EAdS \cite{Sleight:2023ojm,Iacobacci:2024nhw} and therefore the contributions to their conformal block expansion are the same as those in EAdS.

\vskip 4pt
Consider the non-derivative interactions of scalar fields $\phi_i$ in $\mathbb{M}^{d+2}$ of the form
\begin{equation}
    {\cal V}_{12 \ldots n} = g_{12 \ldots n} \phi_1 \phi_2 \ldots \phi_n,
\end{equation}
where $\phi_i$ are real scalar fields with masses $m_i$, $i = 1, 2, \ldots, n$. The $n$-point celestial contact diagram generated by this interaction is proportional to the corresponding contact Witten diagram in EAdS \cite{Iacobacci:2024nhw}:
\begin{align}\label{adstoflatcontact}
{\cal A}^{m_1 \ldots m_n}_{\Delta_1 \ldots \Delta_n}\left(Q_1,\ldots Q_n\right)&=
    -ig \int {\rm d}^{d+2}X\, \prod^n_{i=1}\,G^{(m_i)}_{\Delta_i}\left(X,Q_i\right),\\ &= \,c^{\text{flat-AdS}}_{\Delta_1 \ldots \Delta_n}\left(m_1,\ldots,m_n\right) \nonumber \\& \hspace*{0.5cm} \times  \underbrace{g\int_{{\cal A}_+}{\rm d}^{d+1}{\hat X}\prod^n_{i=1}\, K^{\text{AdS}}_{\Delta}\left(\,  s_{\text{AdS}}({\hat X},Q_i)-i\epsilon\right)}_{\text{Contact Witten diagram in EAdS$_{d+1}$}}, 
\end{align}
where the coefficient 
\begin{multline}\label{adstoflatnpt}
    c^{\text{flat-AdS}}_{\Delta_1 \ldots \Delta_n}\left(m_1,\ldots,m_n\right)=\left(\prod^n_{i=1}c^{\text{dS-AdS}}_{\Delta_i}\right) 2\sin\left(\left(-d+\sum\limits^n_{i=1}\Delta_i\right)\frac{\pi}{2}\right) \\ \times
    R_{\Delta_1 \ldots \Delta_n}\left(m_1,\ldots,m_n\right),
\end{multline}
accounts for the difference in two- and three-point coefficients from AdS to celestial correlators. The integral over the radial direction is encoded by the function
 \begin{equation}\label{RD}
   R_{\Delta_1 \ldots \Delta_n}\left(m_1,\ldots,m_n\right)=\int^{\infty}_{0}{\rm d}R\,R^{d+1}\prod^n_{i=1}\,\mathcal{K}^{(m_i)}_{i\left(\frac{d}{2}-\Delta_i\right)}(R),
\end{equation}
while the sinusoidal factor in \eqref{adstoflatnpt} arises from combining the contributions from regions ${\cal A}_+$ and ${\cal D}_+$, which differ by a phase. 

\vskip 4pt
For massless scalars the radial kernel reduces to a powerlaw \eqref{bubopl}, which significantly simplifies the radial integral \eqref{RD}. In the case that all scalar fields $\phi_i$ are massless $m_i=0$, we have
 \begin{align}\label{Rml}
   R_{\Delta_1 \ldots \Delta_n}\left(0,\ldots,0\right)&=\int^{\infty}_{0}{\rm d}R\,R^{d+1}\prod^n_{i=1}\,R^{\Delta_i-d},\\
   &=2 \pi i\, \delta \left(-(d+2)+\sum\limits^n_{i=1}\left(d-\Delta_i\right)\right),
\end{align}
i.e. the radial contribution reduces to a Dirac delta function enforcing that the shadow scaling dimensions sum up to the dimension of the bulk Minkowski space:
\begin{equation}
    \sum\limits^n_{i=1}\left(d-\Delta_i\right) = d+2.
\end{equation}
All together this gives 
\begin{multline}\label{ctml}
    {\cal A}^{0 \ldots 0}_{\Delta_1 \ldots \Delta_n}\left(Q_1,\ldots Q_n\right) = 2 \pi i\, \delta \left(-(d+2)+\sum\limits^n_{i=1}\left(d-\Delta_i\right)\right) \\ \times \left(\prod^n_{i=1}c^{\text{dS-AdS}}_{\Delta_i}\right) 2\sin\left(\left(-d+\sum\limits^n_{i=1}\Delta_i\right)\frac{\pi}{2}\right) \\ 
    \times  \underbrace{g\int_{{\cal A}_+}{\rm d}^{d+1}{\hat X}\prod^n_{i=1}\, K^{\text{AdS}}_{\Delta}\left(\,  s_{\text{AdS}}({\hat X},Q_i)-i\epsilon\right)}_{\text{Contact Witten diagram in EAdS$_{d+1}$}}.
\end{multline}

\vskip 4pt
It is instructive to consider the case of a single massive scalar $\phi_n$ of mass $m$, where
 \begin{align}
   R_{\Delta_1 \ldots \Delta_n}\left(0,\ldots,m\right)&=\int^{\infty}_{0}{\rm d}R\,R^{d+1}\mathcal{K}^{(m)}_{i\left(\frac{d}{2}-\Delta_n\right)}(R)\left(\prod^{n-1}_{i=1}\,R^{\Delta_i-d}\right).
\end{align}
To evaluate such integrals it is useful to work in Mellin space, where the radial dependence is a power-law:
\begin{multline}\label{MBK}
    \mathcal{K}^{(m)}_{i\left(\frac{d}{2}-\Delta\right)}(R)=\left(\frac{m}{2}\right)^{\frac{d}{2}-\Delta}\frac{1}{\Gamma\left(\frac{d}{2}-\Delta\right)}\\
\times\int_{-i\infty}^{+i\infty}\frac{{\rm d}\omega}{2\pi i}\Gamma\left(\omega+\frac{1}{2}\left(\Delta-\frac{d}{2}\right)\right)\Gamma\left(\omega-\frac{1}{2}\left(\Delta-\frac{d}{2}\right)\right)\left(\frac{m}{2}\right)^{-2\omega}R^{-\frac{d}{2}-2\omega}.
\end{multline}
This reduces the radial integral to a Dirac delta function as in the massless case \eqref{Rml}, giving
 \begin{multline}
   R_{\Delta_1 \ldots \Delta_n}\left(0,\ldots,m\right)=\left(\frac{m}{2}\right)^{\frac{d}{2}-\Delta_n}\frac{1}{\Gamma\left(\frac{d}{2}-\Delta_n\right)}\\
\times\int_{-i\infty}^{+i\infty}\frac{{\rm d}\omega}{2\pi i} 2 \pi i\, \delta \left(-(d+2)+\frac{d}{2}+2\omega+\sum\limits^{n-1}_{i=1}\left(d-\Delta_i\right)\right) \\ \times \Gamma\left(\omega+\frac{1}{2}\left(\Delta_n-\frac{d}{2}\right)\right)\Gamma\left(\omega-\frac{1}{2}\left(\Delta_n-\frac{d}{2}\right)\right)\left(\frac{m}{2}\right)^{-2\omega}.
\end{multline}
The inverse Mellin transform can then be evaluated using the properties Dirac delta function, giving:
 \begin{multline}
   R_{\Delta_1 \ldots \Delta_n}\left(0,\ldots,m\right)=\frac{1}{\Gamma\left(\frac{d}{2}-\Delta_n\right)}\Gamma\left(\frac{d+2}{2}-\sum_{i=1}^{n}\frac{d-\Delta_i}{2}+\frac{d}{2}-\Delta_n\right)\\
\times \Gamma\left(\frac{d+2}{2}-\sum_{i=1}^{n}\frac{d-\Delta_i}{2}\right)\left(\frac{m}{2}\right)^{-2\left(\frac{d+2}{2}-\sum_{i=1}^n\frac{d-\Delta_i}{2}\right)}.
\end{multline}
This recovers the result for all massless scalars \eqref{ctml} in the $m\to 0$ limit via 
\begin{equation}
  \hspace*{-0.65cm} \lim_{m\to 0} \Gamma\left(\frac{d+2}{2}-\sum_{i=1}^{n}\frac{d-\Delta_i}{2}\right)\left(\frac{m}{2}\right)^{-2\left(\frac{d+2}{2}-\sum_{i=1}^n\frac{d-\Delta_i}{2}\right)} = 2\pi i\, \delta\left(\frac{d+2}{2}-\sum_{i=1}^n\frac{d-\Delta_i}{2}\right),
\end{equation}
providing a consistency check of the expression.

\vskip 4pt
Before concluding the discussion on contact diagrams, let us emphasise that since contact diagram contributions to celestial correlators \eqref{ccdefn} are proportional to the corresponding contact Witten diagram in EAdS$_{d+1}$, the conformal families appearing in their conformal block expansion should be the same as those in the corresponding contact Witten diagram. For four-point contact diagrams, these are the double-trace operators \eqref{DTops} arising from the integral over EAdS -- see e.g. \cite{Heemskerk:2009pn,El-Showk:2011yvt,Bekaert:2015tva,Meltzer:2019nbs}.

\subsection{Exchange diagrams and conformal block expansion}
\label{subsec::exchdiagramecbw}

In this section we consider the conformal block expansion \eqref{cbe} of the four-point exchange diagram \eqref{celexch} in the case that all the external scalar fields are massless $m_i=0$ and the exchanged field has generic non-zero mass $m$. The case that the exchanged field is also massless can be obtained from the latter by taking the massless limit $m\to 0$. The latter case has some interesting features which we consider separately in section \ref{subsec::mlexchsc}.

\vskip 4pt
In section \ref{sec::CPWE} we gave the conformal partial wave expansion of the exchange \eqref{cpwecexch}, from which the conformal block decomposition can be obtained with the knowledge of the pole structure of the spectral density \eqref{spectcexch}. Assuming no subtractions are required, this can be written in the form:
\begin{multline}\label{cpwe1cb}
   {\cal A}^{(0,0|m|0,0)}_{\Delta_1 \Delta_2 \Delta_3 \Delta_4}\left(Q_1,Q_2,Q_3,Q_4\right) = \int^{\infty}_{-\infty}\frac{{\rm d}\nu}{2\pi}\,\left[{\bar \rho}^{(0,0|m|0,0)}_{\Delta_1 \Delta_2 \Delta_3 \Delta_4}\left(\nu\right)+{\bar \rho}^{(0,0|m|0,0)}_{\Delta_1 \Delta_2 \Delta_3 \Delta_4}\left(-\nu\right)\right]\\ \times \rho^{\text{AdS}}_{\Delta_1 \Delta_2 \Delta_3 \Delta_4}\left(\nu\right)2\kappa^{34}_{\frac{d}{2}-i\nu,0}{\cal G}^{12,34}_{\frac{d}{2}+i\nu,0}\left(Q_1,Q_2,Q_3,Q_4\right),
\end{multline}
with reduced spectral density
\begin{multline}\label{rhobar}
  \hspace*{-1cm}  {\bar \rho}^{(0,0|m|0,0)}_{\Delta_1 \Delta_2 \Delta_3 \Delta_4}\left(\nu\right)=16 \pi \left(\prod^4_{i=1}c^{\text{dS-AdS}}_{\Delta_i} \right)\sin\left(\tfrac{\pi}{2}\left(-\tfrac{d}{2}-i\nu+\Delta_1+\Delta_2\right)\right)\sin\left(\tfrac{\pi}{2}\left(-\tfrac{d}{2}-i\nu+\Delta_3+\Delta_4\right)\right)\\ \times c^{\text{dS-AdS}}_{\frac{d}{2}+i\nu}\, a^{(0,0|m|0,0)}_{\Delta_1,\Delta_2|-\tfrac{d}{2}+i\nu|\Delta_3,\Delta_4}.
\end{multline}
As explained in section \ref{sec::CPWE} the function $ \rho^{\text{AdS}}_{\Delta_1 \Delta_2 \Delta_3 \Delta_4}\left(\nu\right)$, which arises from the the integration over EAdS, contributes two infinite families of double-trace operators \eqref{DTops} to the conformal block expansion corresponding to the poles:
\begin{subequations}\label{DTmlext}
  \begin{align}
    \frac{d}{2}+i\nu&=\Delta_1+\Delta_2+2n, \qquad n =0, 1, 2, \ldots, \\
    \frac{d}{2}+i\nu&=\Delta_3+\Delta_4+2n, \qquad n =0, 1, 2, \ldots.
\end{align}  
\end{subequations}

\vskip 4pt
The remaining contributions are encoded in the reduced spectral density \eqref{rhobar} and in particular the function \eqref{aexch} which originates from the integral over the radial directions. For external massless scalars (i.e. $m_i=0$ and generic $m$) this is given by (see appendix \ref{app::cpweexch}): 
\begin{multline}\label{aml}
  a^{(0,0|m|0,0)}_{\Delta_1,\Delta_2|-\tfrac{d}{2}+i\nu|\Delta_3,\Delta_4} = \frac{1}{2\pi} \left(\frac{m}{2}\right)^{-4+3d-\sum\limits^4_{i=1}\Delta_i}  \csc\left(\frac{\pi}{2}\left(4-3d+\sum\limits^4_{i=1}\Delta_i\right) \right)\\ \times \sin \left(\tfrac{\pi}{2} \left(2-d+\Delta_1+\Delta_2-\left(\tfrac{d}{2}+i\nu\right)\right)\right)\sin \left(\tfrac{\pi}{2} \left(-\left(\tfrac{d}{2}+i\nu\right)+2-d+\Delta_3+\Delta_4\right)\right)\\ \times \Gamma \left(\tfrac{1}{2} \left(2-d+\Delta_1+\Delta_2-\left(\tfrac{d}{2}\pm i\nu\right)\right)\right)\Gamma \left(\tfrac{1}{2} \left(2-d+\Delta_3+\Delta_4-\left(\tfrac{d}{2}\pm i\nu\right)\right)\right),
\end{multline}
where we introduced the shorthand notation: $\Gamma\left(a \pm b\right)=\Gamma\left(a + b\right)\Gamma\left(a - b\right)$. Notice that the presence of the sinusoidal factors cancels all poles in \eqref{aml} that lie in the lower-half plane. The non-trivial contributions therefore come from its $\nu \to -\nu$ counterpart in \eqref{cpwe1cb}, which gives rise to two families of double-trace-like poles:
\begin{subequations}\label{radialDT}
  \begin{align}
     \frac{d}{2}+i\nu &= \Delta_1+\Delta_2+2n+(2-d), \qquad n = 0, 1, 2, \ldots,\\
     \frac{d}{2}+i\nu &= \Delta_3+\Delta_4+2n+(2-d), \qquad n = 0, 1, 2, \ldots.
\end{align} 
\end{subequations}
These poles encode the exchanged particle in $\mathbb{M}^{d+2}$ and are the analogue of the pole \eqref{STpoleAdS} in the AdS exchange Witten diagram, which encoded the exchanged single-particle state in AdS. I.e. They arise from a single Poincar\'e block for a scalar of mass $m$ after radial reduction of the theory on the co-dimension 1 extended unit hyperboloid.

\vskip 4pt
Let us note that the same contributions can be obtained by starting from the asymmetric expression \eqref{cpwecexch} for the conformal partial wave expansion, pulling the integration contour down for the term involving ${\cal G}^{12,34}_{\frac{d}{2}+i\nu,0}$ and up for the term involving ${\cal G}^{12,34}_{\frac{d}{2}-i\nu,0}$. In this case one does not confront the possibility of having to perform subtractions to make the spectral integral well defined, which we discuss further below, making the asymmetric representation more convenient in such cases.

\vskip 4pt
The contributions to the conformal block expansion of the celestial exchange obtained in this section are in perfect agreement with those obtained in \cite{Pacifico:2024dyo} from the Mellin amplitude representation of the exchange diagram \eqref{celexch}. In appendix \ref{app::spectralintegral} we show that by evaluating the spectral integral in the conformal partial wave expansion \eqref{Fexch2cb} one recovers its Mellin amplitude representation derived in \cite{Pacifico:2024dyo}. This provides a non-trivial check of approach presented in this work and of the analytic continuations to EAdS derived in \cite{Iacobacci:2024nhw}.

\paragraph{Subtractions.} The expression \eqref{cpwe1cb} for the conformal partial wave expansion is valid so long as both terms share the same integration contour. This is possible when none of the poles in the reduced spectral function \eqref{rhobar} cross the real axis. From the explicit expression \eqref{aml}, assuming that $\Delta_i=\frac{d}{2}+i\nu_i$ belong to principal series representations of the Lorentz group (i.e. $\nu_i \in \mathbb{R}$), this holds for $d < 4$. For $d \geq 4$ one must add/subtract the residues of poles that are not captured by both contours. Alternatively one can work with the asymmetric expression \eqref{cpwecexch} for the conformal partial wave expansion, where this issue does not arise.

\subsection{Special case: Massless exchanged scalar}
\label{subsec::mlexchsc}

Let us now consider the case that the exchanged field is also massless i.e. $m=0$. Like for the contact diagrams, when all fields are massless a constraint emerges on the external scaling dimensions, which can be seen from the massless limit of the massive result \eqref{cpwe1cb} via:
\begin{multline}\label{DDfunction}
\lim_{m\to0}\left[m^{-4+3d-\sum\limits^4_{i=1}\Delta_i}\Gamma\left(\frac{1}{2}\left(4-3d+\sum\limits^4_{i=1}\Delta_i\right)\right)\right]=2\pi i\,\delta\left(\frac{1}{2}\left(4-3d+\sum\limits^4_{i=1}\Delta_i\right)\right).
\end{multline}
This leads to some significant simplifications with respect to the massive case.

\vskip 4pt
Let us first note that the symmetric form \eqref{cpwecexch} for the conformal partial wave expansion is no longer valid for any $d$, since the constraint \eqref{DDfunction} among the scaling dimensions implies that there are poles in $\nu$ that cross the real axis for any $d$. In fact, without performing the required subtractions, combining the constraint \eqref{DDfunction} with the expression \eqref{cpwe1cb} for the conformal partial wave expansion would naively give a vanishing result: On the support of the Dirac delta function \eqref{DDfunction}, the term in the square bracket in \eqref{cpwe1cb} is vanishing for integer dimensions $d$. 

\vskip 4pt
In this case it is more convenient to work with the asymmetric form \eqref{Fexch2cb} for the conformal partial wave expansion \eqref{Fexch2cb}, which generates the conformal block expansion by pulling the integration contour down for the term involving ${\cal G}^{12,34}_{\frac{d}{2}+i\nu,0}$ and up for the term involving ${\cal G}^{12,34}_{\frac{d}{2}-i\nu,0}$. Applying the Dirac delta function constraint \eqref{DDfunction}, from the term involving ${\cal G}^{12,34}_{\frac{d}{2}+i\nu,0}$ there are no non-trivial contributions from poles in the lower-half plane. From the term involving ${\cal G}^{12,34}_{\frac{d}{2}-i\nu,0}$, the contributions in the upper-half plane come from the factors:\footnote{$\left(z\right)_n = \frac{\Gamma\left(z+n\right)}{\Gamma\left(z\right)}$ is the rising Pochhammer symbol.}
\begin{equation}\label{rhomllower}
   \rho^{(0,0|0|0,0)}_{\Delta_1 \Delta_2 \Delta_3 \Delta_4}\left(\nu\right) \sim \frac{\left(1+\tfrac{1}{2}\left(\tfrac{d}{2}+i\nu\right)-\tfrac{1}{2}\left(\Delta_1+\Delta_2\right)\right)_{d-3}}{\frac{d}{2}-1+\frac{1}{2} \left(\tfrac{d}{2}+i\nu\right)-\frac{1}{2} (\Delta_1+\Delta_2)}\frac{\left(1+\tfrac{1}{2}\left(\tfrac{d}{2}+i\nu\right)-\tfrac{1}{2}\left(\Delta_3+\Delta_4\right)\right)_{d-3}}{\frac{d}{2}-1+\frac{1}{2} \left(\tfrac{d}{2}+i\nu\right)-\frac{1}{2} (\Delta_3+\Delta_4)}. 
\end{equation}
We see that, for any integer $d$, there are a finite number of poles in the upper-half plane,\footnote{These would correspond to the finite number of subtractions required to obtain a well-defined symmetric form \eqref{cpwe1cb} for the conformal partial wave expansion.} so the exchange diagram in this case is given exactly by a finite sum of conformal blocks. For $d \geq 3$ these are:\footnote{For ease of presentation in \eqref{mlcbgend} we have left the Dirac delta function constraint \eqref{DDfunction} on the scaling dimensions implicit.}
\begin{multline}
\hspace*{-1cm}    {\cal A}^{(0,0|0|0,0)}_{\Delta_1 \Delta_2 \Delta_3 \Delta_4}\left(Q_i\right)= 2 \sin \left(\frac{\pi  d}{2}\right) \Gamma \left(\frac{d-2}{2}\right)^2 \\ \times \Gamma \left(-\tfrac{d}{2}+\Delta_1+1\right) \Gamma \left(-\tfrac{d}{2}+\Delta_2+1\right) \Gamma (d-\Delta_3-1) \Gamma (d-\Delta_4-1)\\
   \times\frac{\sin (\pi  (d-\Delta_1-\Delta_2)) \Gamma \left(\tfrac{3 d}{2}-\Delta_1-\Delta_2-2\right) \Gamma (2 d-\Delta_1-\Delta_2-3)}{-d+\Delta_1+\Delta_2+1}\ {\cal G}^{12,34}_{\Delta_1+\Delta_2+2-d,0}\left(Q_i\right)\\
    +(\Delta_1,\ \Delta_2\ \leftrightarrow\ \Delta_3,\ \Delta_4),\label{mlcbgend}
\end{multline}
which correspond to contributions from radial two double-trace operators \eqref{radialDT} with $n=0$:
\begin{equation}\label{odddt}
    \Delta_{1}+\Delta_2+2-d, \qquad \Delta_{3}+\Delta_4+2-d.
\end{equation}
Interestingly, the above result can be rewritten in terms of a single Conformal Partial Wave encoding both contributions, as presented in \cite{Pacifico:2024dyo}. This is by virtue of the Dirac delta function constraint \eqref{DDfunction}, which implies that the scaling dimensions \eqref{odddt} are shadow of one another.

\vskip 4pt
 For even $d \geq 4$ the poles displayed in \eqref{rhomllower} all cancel and the corresponding celestial exchange diagram seems to be vanishing (at least for non-null separated points $Q_i \cdot Q_j \ne 0$). Indeed, the coefficient of each conformal block in the expression \eqref{mlcbgend} is vanishing for even integer $d > 2$. For $d=2$ instead we have
\begin{equation}
   \rho^{(0,0|0|0,0)}_{\Delta_1 \Delta_2 \Delta_3 \Delta_4}\left(\nu\right) \sim \frac{1}{\left(-\left(\frac{d}{2}+i\nu\right) +\Delta_1+\Delta_2)\right)^2\left(-\left(\frac{d}{2}+i\nu\right) +\Delta_3+\Delta_4)\right)^2},
\end{equation}
with double poles corresponding to the two double-trace scaling dimensions: 
\begin{equation}\label{mldtdims}
    \Delta_1+\Delta_2, \qquad \Delta_3+\Delta_4.
\end{equation}
 The exchange diagram in this case is a sum of the two conformal blocks and their derivatives:
\begin{multline}\label{exchd2}
    \mathcal{A}^{0,0|0|0,0}_{\Delta_1\Delta_2\Delta_3\Delta_4}=\delta(\Delta_1+\Delta_2+\Delta_3+\Delta_4-2)\Bigg[\frac{f'(\Delta_1+\Delta_2)}{4 (\Delta_1+\Delta_2-1)^2}+\frac{f'(\Delta_3+\Delta_4)}{4 (\Delta_3+\Delta_4-1)^2}\\+\frac{f(\Delta_1+\Delta_2)}{4 (\Delta_1+\Delta_2-1)^3}-\frac{f(\Delta_3+\Delta_4)}{4 (\Delta_3+\Delta_4-1)^3}\Bigg],
\end{multline}
where for simplicity we defined $f(\Delta)=\mathcal{N}_\Delta {\cal G}^{12,34}_{\Delta,0}$, with
\begin{align}
    \mathcal{N}_{\frac{d}{2}+i\nu}&=\Gamma (1-\Delta_1) \Gamma (1-\Delta_2) \Gamma (1-\Delta_3) \Gamma (1-\Delta_4)\\
    &\hspace*{-0.8cm}\times\frac{\Gamma (1+i \nu) \Gamma \left(\frac{\Delta_1-\Delta_2-i \nu +1}{2}\right) \Gamma \left(\frac{-\Delta_1+\Delta_2-i \nu +1}{2}\right) \Gamma \left(\frac{\Delta_3-\Delta_4-i \nu +1}{2}\right) \Gamma \left(\frac{-\Delta_3+\Delta_4-i \nu +1}{2}\right)}{16 \pi ^5 \Gamma (1-i \nu )}\,.\nonumber
\end{align}
Conformal blocks in $d=2$ have a relatively simple closed form expression in terms of Gauss hypergeometric functions \cite{Ferrara:1974ny,Dolan:2000ut}. Combined with the expression \eqref{exchd2}, this gives a relatively simple closed form expression for the massless celestial exchange in $d=2$.

\vskip 4pt
In summary, see that when the exchanged field is massless the contributions \eqref{radialDT} to the conformal block expansion from the infinite multiplet of operators present in the massive case collapses to contributions from the first operator in each series. This shortening of the multiplet can be understood conceptually as arising from the one-to-one correspondence between massless fields in Minkowski space and massive fields on the co-dimension 1 hyperboloid \cite{Fronsdal:1978vb}, which therefore extends to the exchange of massless spinning fields in Minkowski space as well \cite{toappear}.

\section{Outlook: Towards a non-perturbative celestial bootstrap}
\label{sec::outlook}

In this work we have shown that perturbative celestial correlators \eqref{ccdefnint} for scalar field theories in $(d+2)$-dimensional Minkowski space admit a conformal partial wave expansion
\begin{multline}\label{cpweconcl}
    \langle {\cal O}_{\Delta_1}\left(Q_1\right){\cal O}_{\Delta_2}\left(Q_2\right){\cal O}_{\Delta_3}\left(Q_3\right){\cal O}_{\Delta_4}\left(Q_4\right) \rangle = \sum\limits_{J}\int^{\infty}_{-\infty}\frac{{\rm d}\nu}{2\pi}\,\rho^{\Delta_1,\Delta_2,\Delta_3,\Delta_4}_{J}\left(\nu\right)\\ \times {\cal F}^{\Delta_1,\Delta_2,\Delta_3,\Delta_4}_{\nu,J}\left(Q_1,Q_2,Q_3,Q_4\right),
\end{multline}
with meromorphic spectral density $\rho_{J}^{\Delta_1,\Delta_2,\Delta_3,\Delta_4}\left(\nu\right)$. We have also presented tools to determine this expansion at any order in perturbation theory, by extending the AdS harmonic function decomposition \cite{Moschella:2007zza,Penedones:2007ns,Costa:2014kfa} to the Minkowski Feynman propagator \eqref{bubuharm0}. One can now use the same argument of \cite{Sleight:2020obc,Sleight:2021plv} to conclude that celestial correlators admit a meromorphic spectral decomposition, at least at the perturbative level.

\vskip 4pt
An important open question is whether celestial correlators admit a conformal partial wave expansion \eqref{cpweconcl} with a meromorphic spectral function also at the non-perturbative level.\footnote{For boundary correlators in AdS space this follows from the state-operator correspondence \cite{Paulos:2016fap}.} Assuming this is the case, as suggested by perturbation theory to all orders, one can in principle set up a non-perturbative bootstrap for celestial correlators \eqref{ccdefnint} similar to the AdS case, where unitarity\footnote{I.e. positivity of the norm for states carrying unitary irreducible representations of the Euclidean conformal group $SO(1,d+1)$ acting on the Hilbert space of the theory. Note here the difference with respect to Lorentzian AdS, where the positivity of the norm must be imposed before Wick rotation. After Wick rotation this translates into reflection positivity rather then unitarity for the Euclidean theory.} of the Euclidean conformal field theory on the celestial sphere imposes positivity of the spectral density:
\begin{equation}
    \rho_{J}^{\Delta_1,\Delta_2,\Delta_1^*,\Delta_2^*}\left(\nu\right) \geq 0.
\end{equation}
A condition of this type is quite common when decomposing a positive norm state $|\psi\rangle\in\mathcal{H}$ into a complete basis of orthonormal states $|\psi_\nu\rangle$:
\begin{align}
    1=\sum\int {\rm d}\nu |\psi_\nu\rangle\langle\psi_\nu|,
\end{align}
where $\nu$ can be both a continuous or discrete set of quantum numbers. One can then write the Plancherel formula:
\begin{align}\label{Planch}
    \langle\psi|\psi\rangle=\sum\int {\rm d}\nu \langle\psi|\psi_\nu\rangle\langle\psi_\nu|\psi\rangle=\sum\int {\rm d}\nu|\langle\psi_\nu|\psi\rangle|^2,
\end{align}
recovering the positivity condition $|\langle\psi_\nu|\psi\rangle|^2\geq0$. In the case at hand, one considers the state\footnote{Note that in Euclidean signature the order of the operators is irrelevant so that the function $f(Q_1,Q_2)$ can be chosen to be symmetric under the exchange of the two variables.}
\begin{align}
    |\psi\rangle=\int {\rm d}Q_i\,f(Q_1,Q_2)\, O_{\Delta_1}(Q_1)O_{\Delta_2}(Q_2)|0\rangle,
\end{align}
where $f(Q_1,Q_2)\in\mathcal{S}(\mathbb{R}^{2d})$ is a Schwartz function and evaluate the norm of this state inserting as complete basis the principal series representation appearing in the Plancherel formula \eqref{Planch}:\footnote{Below we do not consider discrete series contribution which could be there in the Plancherel formula but do not appear in our analysis.}
\begin{align}\label{Plancherel}
    1=\sum_J\int \frac{{\rm d}\nu}{2\pi}\int {\rm d}Q\,|\psi_{\nu,J}(Q)\rangle\langle\psi_{\nu,J}(Q)|+\ldots\,.
\end{align}
Here $|\psi_{\nu,J}(Q)\rangle$ admits a convenient plane wave representation given in terms of the bulk-to-boundary propagator, which in our notation (e.g. for $J=0$) reads:
\begin{align}
    \langle X|\psi_{\nu,0}(Q)\rangle\sim\frac{1}{(-2X\cdot Q)^{\frac{d}{2}+i\nu}}\,.
\end{align}
Using the above Plancherel formula one can then arrives to the inequality:
\begin{multline}
    \int {\rm d}Q_i\,f(Q_1,Q_2)\bar{f}(Q_3,Q_4)\\ \times \int {\rm d} Q\,\langle O_2^\dagger(Q_4)O_1^\dagger(Q_3)|\psi_{\nu,J}(Q)\rangle\langle \psi_{\nu,J}(Q)|O_1(Q_1)O_2(Q_2)\rangle\geq0,
\end{multline}
valid for any choice of the function $f$. Conformal symmetry now implies that the latter integrand is proportional to the conformal partial wave since it has the same quantum numbers up to an overall coefficient which defines the spectral function:
\begin{multline}
    \langle O_2^\dagger(Q_4)O_1^\dagger(Q_3)|\psi_{\nu,J}(Q)\rangle\langle \psi_{\nu,J}(Q)|O_1(Q_1)O_2(Q_2)\rangle \\ =\rho_{J}^{\Delta_1^*,\Delta_2^*,\Delta_1,\Delta_2}(\nu)\mathcal{F}_{\nu,J}^{\Delta_1^*\Delta_2^*\Delta_1\Delta_2}(Q_3,Q_4,Q_1,Q_2)\,.
\end{multline}
One can now use the following inequality:\footnote{This inequality is manifest in momentum space. In fact in momentum space the CPW is completely factorised and is the product of the two 3pt conformal correlators. For complex conjugate dimensions on the principal series this is manifestly a modulus squared. See e.g. \cite{Sleight:2021iix} eq. (2.2).}
\begin{align}
    \int {\rm d}Q_i\, f(Q_1,Q_2)\bar{f}(Q_3,Q_4)\mathcal{F}_{\nu,J}^{\Delta_1^*\Delta_2^*\Delta_1\Delta_2}(Q_3,Q_4,Q_1,Q_2)\geq0\,,
\end{align}
which can be easily verified since it has the form of a squared norm, to show that
\begin{align}
    \rho_J^{\Delta_1^*\Delta_2^*\Delta_1\Delta_2}(\nu)\geq0\,.
\end{align}
Therefore positivity of the norm of the Hilbert space states (a.k.a. unitarity) easily translates into positivity of the spectral density as usual in Quantum Mechanics. The main difference with respect to Lorentzian AdS is that in AdS the spectrum of unitary representations appearing in the completeness relation \eqref{Plancherel} is discrete (the unitary irreducible representations are lowest weight) and gives rise to an expansion of positive OPE coefficients squared. In flat space (like in dS), the spectrum is continuous and the correct complete basis given by Principal series (plus potential discrete contributions\footnote{These are expected to play a role for gauge fields and gravity.}).

\vskip 4pt
In the more general case the situation is more involved since the external legs can be arbitrary principal series representation which all belong to the spectrum of the theory. In this case $\Delta_i$ can be all different but one can still consider the positive norm state
\begin{align}
    |\psi\rangle=\int {\rm d}Q_i\,f(Q_1,Q_2) O_1(Q_1)O_2(Q_2)|0\rangle+\int {\rm d}Q_i\,g(Q_1,Q_2)\, O_3(Q_1)O_4(Q_2)|0\rangle,
\end{align} for arbitrary test functions $f$ and $g$, leading to a positivity condition for the matrix:
\begin{align}
\begin{pmatrix}
\rho_J^{\Delta_1^*,\Delta_2^*,\Delta_1,\Delta_2}&\rho_J^{\Delta_1^*,\Delta_2^*,\Delta_3,\Delta_4}\\
\rho_J^{\Delta_3^*,\Delta_4^*,\Delta_1,\Delta_2}&\rho_J^{\Delta_3^*,\Delta_4^*,\Delta_3,\Delta_4}
\end{pmatrix}\geq0\,.
\end{align}

\vskip 4pt Let us finally comment that for celestial correlators an additional subtlety is present with respect to the AdS and dS cases: The above positivity conditions are only necessary conditions for unitarity, since one must ensure that the operator spectrum organises in terms of unitary \emph{Poincar\'e} representations.

\section*{Acknowledgments}

We thank Sébastien Malherbe, Kevin Nguyen, Michel Pannier and Paolo Pergola for discussions. This research was supported by the European Union (ERC grant ``HoloBoot'', project number 101125112),\footnote{Views and opinions expressed are however those of the author(s) only and do not necessarily reflect those of the European Union or the European Research Council. Neither the European Union nor the granting authority can be held responsible for them.} by the MUR-PRIN grant No. PRIN2022BP52A (European Union - Next Generation EU) and by the INFN initiative STEFI. The research of FP was partially supported by an Erasmus+ scholarship. The research of CS was partially supported by the STFC grant ST/T000708/1 of the Mathematical and
Theoretical Particle Physics group at Durham University.

\newpage

\begin{appendix}

\section{AdS harmonic function decomposition of the Feynman propagator}
\label{app::FeynAdS}

In this appendix we give further details on the AdS harmonic function decomposition \eqref{bubuharm} of the Minkowski Feynman propagator. 

\vskip 4pt
For concreteness let us take the two bulk points to lie in the region ${\cal D}_+$. The starting point is the decomposition \eqref{Feynasbubuads} of the Feynman propagator in terms of bulk-to-bulk propagators in EAdS (via analytic continuation):
\begin{equation}
    G^{(m)}_T(X_{{\cal D}_+},Y_{{\cal D}_+})
   \to \int_{-\infty}^{+\infty}\frac{{\rm d}\nu}{2\pi}\,\rho_\nu^{(m)}(R_1,R_2)e^{-(\frac{d}{2}+i\nu)\pi i}c^{\text{dS-AdS}}_{\frac{d}{2}+i\nu}G^{\text{AdS}}_{\frac{d}{2}+i\nu}({\hat X}_{{\cal A}_+},{\hat Y}_{{\cal A}_+}),
\end{equation}
where 
\begin{align}
\rho^{\left(m\right)}_\nu\left(R_1,R_2\right)&=\mathcal{K}_\nu^{(m)}(R_1)\mathcal{K}_{-\nu}^{(m)}(R_2\,).
\end{align}
This is symmetric under $\nu \to -\nu$, so that equivalently we can write:
\begin{multline}\label{appfeynsymm}
    G^{(m)}_T(X_{{\cal D}_+},Y_{{\cal D}_+})
   \to \int_{-\infty}^{+\infty}\frac{{\rm d}\nu}{2\pi}\,\frac{1}{2}\rho_\nu^{(m)}(R_1,R_2)\left[e^{-(\frac{d}{2}+i\nu)\pi i}c^{\text{dS-AdS}}_{\frac{d}{2}+i\nu}G^{\text{AdS}}_{\frac{d}{2}+i\nu}({\hat X}_{{\cal A}_+},{\hat Y}_{{\cal A}_+})\right.\\\left.+\nu\to-\nu\right].
\end{multline}
Being symmetric in $\nu$, this has poles both in the upper- and lower-half $\nu$-plane. However, since in \eqref{appfeynsymm} it is now multiplying an even function of $\nu$, this can be replaced with an asymmetric function of $\nu$ which has poles that are either all in the lower-half plane or all in the upper half plane. To this end, following section 5.2 of \cite{Iacobacci:2024nhw} it is useful to employ the Mellin-Barnes representation \eqref{MBK} for the radial kernel:
\begin{multline}
    \rho^{\left(m\right)}_\nu\left(R_1,R_2\right)=\frac{1}{\Gamma\left(i\nu\right)\Gamma\left(-i\nu\right)}\int_{-i\infty}^{+i\infty}\frac{{\rm d}\omega}{2\pi i}\frac{{\rm d}{\bar \omega}}{2\pi i}\Gamma\left(\omega+\frac{i\nu}{2}\right)\Gamma\left(\omega-\frac{i\nu}{2}\right)\\ \times \Gamma\left({\bar \omega}+\frac{i\nu}{2}\right)\Gamma\left({\bar \omega}-\frac{i\nu}{2}\right)\left(\frac{m}{2}\right)^{-2(\omega+{\bar \omega})}R^{-\frac{d}{2}-2\omega}_1R^{-\frac{d}{2}-2{\bar \omega}}_2.
\end{multline}
Consider now the identity:
\begin{multline}\label{TrigID}
\frac{i\nu}{\pi}\,\Gamma(i\nu)\Gamma(-i\nu)\csc \left(\pi (\omega+{\bar \omega})\right) \sin \left(\pi\left(\omega-\frac{i \nu}{2} \right)\right) \sin \left(\pi\left({\bar \omega}-\frac{i \nu}{2} \right)\right)\\+(\nu \to -\nu)=1\,.
\end{multline}
The zeros of the sine factors in the left/right term precisely cancel the poles below/above the integration contour. Inserting into \eqref{appfeynsymm} and changing integration variable for one of the two terms gives
\begin{multline}
    G^{(m)}_T(X_{{\cal D}_+},Y_{{\cal D}_+})
   \to \int_{-\infty}^{+\infty}\frac{{\rm d}\nu}{2\pi}\,2i\nu\,g_{-\tfrac{d}{2}+i\nu}^{(m)}(R_1,R_2) \left[e^{-(\frac{d}{2}+i\nu)\pi i}c^{\text{dS-AdS}}_{\frac{d}{2}+i\nu}G^{\text{AdS}}_{\frac{d}{2}+i\nu}({\hat X}_{{\cal A}_+},{\hat Y}_{{\cal A}_+})\right.\\\left.+\left(\nu \to -\nu\right)\right],
\end{multline}
where 
\begin{multline}\label{appgMB}
    g^{(m)}_{-\tfrac{d}{2}+i\nu}(R_1,R_2) = \frac{1}{2\pi} \int^{+i\infty}_{-i\infty}\frac{{\rm d}\omega}{2\pi i}\frac{{\rm d}{\bar \omega}}{2\pi i} \csc\left(\pi\left(\omega+{\bar \omega}\right) \right) \sin \left(\pi \left(\omega-\tfrac{i\nu}{2}\right)\right)\sin \left(\pi \left({\bar \omega}-\tfrac{i\nu}{2}\right)\right)\\ \times\Gamma\left(\omega+\tfrac{i\nu}{2}\right)\Gamma\left(\omega-\tfrac{i\nu}{2}\right)\Gamma\left(\bar{\omega}+\tfrac{i\nu}{2}\right)\Gamma\left(\bar{\omega}-\tfrac{i\nu}{2}\right) \left(\frac{m}{2}\right)^{-2\left(\omega+{\bar \omega}\right)} \\ \times R^{-\frac{d}{2}-2\omega}_1 R^{-\frac{d}{2}-2{\bar \omega}}_2.
\end{multline}
It is interesting to note that functionally this is the same as a momentum space bulk-to-bulk propagator in EAdS$_{1-d}$ for a scalar field with scaling dimension $-\tfrac{d}{2}+i\nu$, where the mass $m$ plays the role of the modulus of the boundary momentum and $R_1, R_2$ the bulk Poincar\'e coordinates (\emph{cf.} equation (2.9) of \cite{Sleight:2020obc}). Equivalently, this can be written in the more familiar form \cite{Liu:1998ty}:
\begin{multline}
    g_{-\tfrac{d}{2}+i\nu}^{(m)}(R_1,R_2)= R^{-\frac{d}{2}}_1R^{-\frac{d}{2}}_2\Big[\theta(R_2-R_1) K_{i\nu}\left(m R_1\right)I_{i\nu}\left(m R_2\right)\\+\theta(R_1-R_2) K_{i\nu}\left(m R_2\right)I_{i\nu}\left(m R_1\right)\Big].
\end{multline}

\section{Conformal partial wave expansion of the exchange diagram}
\label{app::cpweexch}

In this appendix we give further details on the conformal partial wave expansion \eqref{cpwecexch} of the celestial exchange diagram \eqref{celexch}. Applying the Feynman rules, the latter reads:
\begin{multline}
  \hspace*{-0.5cm} {\cal A}^{(m_1,m_2|m|m_3,m_4)}_{\Delta_1 \Delta_2\Delta_3 \Delta_4}\left(Q_1,\ldots,Q_4\right)=\left(-ig_{12}\right)\left(-ig_{34}\right) \int {\rm d}^{d+2}X\,\int {\rm d}^{d+2}Y\,G^{(m_1)}_{\Delta_1}\left(X,Q_1\right)G^{(m_2)}_{\Delta_2}\left(X,Q_2\right)\\ \times G^{\left(m\right)}_{T}\left(X,Y\right)G^{(m_3)}_{\Delta_3}\left(Y,Q_3\right)G^{(m_4)}_{\Delta_4}\left(Y,Q_4\right). 
\end{multline}
As reviewed in section \ref{sec::CCF}, the combination of regions ${\cal A}_-$ and ${\cal D}_-$ in the hyperbolic slicing of Minkowski space give a vanishing contribution to celestial correlators \eqref{ccdefn}. We therefore only need to consider contributions from regions ${\cal A}_+$ and ${\cal D}_+$:
\begin{equation}\label{appexchtot}
{\cal A}^{(m_1,m_2|m|m_3,m_4)}_{\Delta_1 \Delta_2\Delta_3 \Delta_4}\left(Q_1,\ldots,Q_4\right)=I_{{\cal D}_+{\cal D}_+}+I_{{\cal D}_+{\cal A}_+}+I_{{\cal A}_+{\cal D}_+}+I_{{\cal A}_+{\cal A}_+},
\end{equation}
where 
\begin{multline}
    I_{\bullet \bullet}=\left(-ig_{12}\right)\left(-ig_{34}\right) \int_{\bullet} {\rm d}^{d+2}X\,\int_{\bullet} {\rm d}^{d+2}Y\,G^{(m_1)}_{\Delta_1}\left(X,Q_1\right)G^{(m_2)}_{\Delta_2}\left(X,Q_2\right)\\ \times G^{\left(m\right)}_{T}\left(X,Y\right)G^{(m_3)}_{\Delta_3}\left(Y,Q_3\right)G^{(m_4)}_{\Delta_4}\left(Y,Q_4\right).
\end{multline}
By analytically continuing region ${\cal D}_+$ to ${\cal A}_+$ according to \eqref{dsadscont}, inserting the AdS harmonic function decomposition \eqref{bubuharm} of the Feynman propagator and the expressions \eqref{bubotoads} for the celestial bulk-to-boundary propagators, we have:
\begin{multline}
 \hspace*{-0.65cm} I_{\bullet \bullet}= \int^{\infty}_{-\infty}\frac{{\rm d}\nu}{2\pi}\,4\pi\, c^{\text{dS-AdS}}_{\frac{d}{2}+i\nu} \left(\prod^4_{i=1}c^{\text{dS-AdS}}_{\Delta_i} \right)\rho^{\bullet \bullet}\left(\nu\right)\rho^{\text{AdS}}_{\Delta_1 \Delta_2 \Delta_3 \Delta_4}\left(\nu\right) {\cal F}^{12,34}_{\nu,0}\left(Q_1,Q_2,Q_3,Q_4\right),
\end{multline}
where 
\begin{subequations}
    \begin{align}\nonumber
        \rho^{{\cal D}_+{\cal D}_+}\left(\nu\right)&=e^{-(\frac{d}{2}-i\nu)\pi i} e^{-(\Delta_1+\Delta_2+\Delta_3+\Delta_4)\frac{\pi i}{2}}\int^\infty_0 {\rm d}R_1 {\rm d}R_2 \,R^{d+1}_1 R^{d+1}_2\,\mathcal{K}^{(m_1)}_{i\left(\frac{d}{2}-\Delta_1\right)}(R_1)\mathcal{K}^{(m_2)}_{i\left(\frac{d}{2}-\Delta_2\right)}(R_1)\\ & \hspace*{4cm}\times g^{(m)}_{-\tfrac{d}{2}+i\nu}(R_1,R_2)\mathcal{K}^{(m_3)}_{i\left(\frac{d}{2}-\Delta_3\right)}(R_2)\mathcal{K}^{(m_4)}_{i\left(\frac{d}{2}-\Delta_4\right)}(R_2),\\ \nonumber
        \rho^{{\cal A}_+{\cal D}_+}\left(\nu\right)&= e^{(\Delta_1+\Delta_2-\Delta_3-\Delta_4)\frac{\pi i}{2}}\int^\infty_0 {\rm d}R_1 {\rm d}R_2 \,R^{d+1}_1 R^{d+1}_2\,\mathcal{K}^{(m_1)}_{i\left(\frac{d}{2}-\Delta_1\right)}(R_1e^{\frac{\pi i}{2}})\mathcal{K}^{(m_2)}_{i\left(\frac{d}{2}-\Delta_2\right)}(R_1e^{\frac{\pi i}{2}})\\ & \hspace*{3cm}\times g^{(m)}_{-\tfrac{d}{2}+i\nu}(R_1 e^{\frac{\pi i}{2}},R_2)\mathcal{K}^{(m_3)}_{i\left(\frac{d}{2}-\Delta_3\right)}(R_2)\mathcal{K}^{(m_4)}_{i\left(\frac{d}{2}-\Delta_4\right)}(R_2),\\ \nonumber
        \rho^{{\cal D}_+{\cal A}_+}\left(\nu\right)&= e^{(\Delta_3+\Delta_4-\Delta_1-\Delta_2)\frac{\pi i}{2}}\int^\infty_0 {\rm d}R_1 {\rm d}R_2 \,R^{d+1}_1 R^{d+1}_2\,\mathcal{K}^{(m_1)}_{i\left(\frac{d}{2}-\Delta_1\right)}(R_1)\mathcal{K}^{(m_2)}_{i\left(\frac{d}{2}-\Delta_2\right)}(R_1)\\ & \hspace*{2.5cm}\times g^{(m)}_{-\tfrac{d}{2}+i\nu}(R_1,R_2e^{\frac{\pi i}{2}})\mathcal{K}^{(m_3)}_{i\left(\frac{d}{2}-\Delta_3\right)}(R_2e^{\frac{\pi i}{2}})\mathcal{K}^{(m_4)}_{i\left(\frac{d}{2}-\Delta_4\right)}(R_2e^{\frac{\pi i}{2}}),\\ \nonumber
        \rho^{{\cal A}_+{\cal A}_+}\left(\nu\right)&= e^{(\frac{d}{2}-i\nu)\pi i} e^{(\Delta_1+\Delta_2+\Delta_3+\Delta_4)\frac{\pi i}{2}}\int^\infty_0 {\rm d}R_1 {\rm d}R_2 \, R^{d+1}_1 R^{d+1}_2\,\mathcal{K}^{(m_1)}_{i\left(\frac{d}{2}-\Delta_1\right)}(R_1e^{\frac{\pi i}{2}})\mathcal{K}^{(m_2)}_{i\left(\frac{d}{2}-\Delta_2\right)}(R_1e^{\frac{\pi i}{2}})\\ & \hspace*{1.5cm}\times g^{(m)}_{-\tfrac{d}{2}+i\nu}(R_1e^{\frac{\pi i}{2}},R_2e^{\frac{\pi i}{2}})\mathcal{K}^{(m_3)}_{i\left(\frac{d}{2}-\Delta_3\right)}(R_2e^{\frac{\pi i}{2}})\mathcal{K}^{(m_4)}_{i\left(\frac{d}{2}-\Delta_4\right)}(R_2e^{\frac{\pi i}{2}}). \label{agenm}
    \end{align}
\end{subequations}
These are all (up to a phase) given by the radial integral:
\begin{multline}\label{aapp}
  a^{(m_1,m_2|m|m_3,m_4)}_{\Delta_1,\Delta_2|-\tfrac{d}{2}+i\nu|\Delta_3,\Delta_4} = \int^\infty_0 {\rm d}R_1 {\rm d}R_2 R^{d+1}_1 R^{d+1}_2\,\mathcal{K}^{(m_1)}_{i\left(\frac{d}{2}-\Delta_1\right)}(R_1)\mathcal{K}^{(m_2)}_{i\left(\frac{d}{2}-\Delta_2\right)}(R_1)\\ \times g^{(m)}_{-\tfrac{d}{2}+i\nu}(R_1,R_2)\mathcal{K}^{(m_3)}_{i\left(\frac{d}{2}-\Delta_3\right)}(R_2)\mathcal{K}^{(m_4)}_{i\left(\frac{d}{2}-\Delta_4\right)}(R_2).
\end{multline}
 In the full expression for the exchange \eqref{appexchtot} thee phases combine to give sinusoidal factors:
\begin{multline}
   {\cal A}^{(m_1,m_2|m|m_3,m_4)}_{\Delta_1 \Delta_2 \Delta_3 \Delta_4}\left(Q_1,Q_2,Q_3,Q_4\right)\\ = \int^{\infty}_{-\infty}\frac{{\rm d}\nu}{2\pi}\,\rho^{(m_1,m_2|m|m_3,m_4)}_{\Delta_1 \Delta_2 \Delta_3 \Delta_4}\left(\nu\right){\cal F}^{12,34}_{\nu,0}\left(Q_1,Q_2,Q_3,Q_4\right),
\end{multline}
with
\begin{multline}
  \hspace*{-1cm}  \rho^{(m_1,m_2|m|m_3,m_4)}_{\Delta_1 \Delta_2 \Delta_3 \Delta_4}\left(\nu\right)=16 \pi \left(\prod^4_{i=1}c^{\text{dS-AdS}}_{\Delta_i} \right)\sin\left(\tfrac{\pi}{2}\left(-\tfrac{d}{2}-i\nu+\Delta_1+\Delta_2\right)\right)\sin\left(\tfrac{\pi}{2}\left(-\tfrac{d}{2}-i\nu+\Delta_3+\Delta_4\right)\right)\\ \times c^{\text{dS-AdS}}_{\frac{d}{2}+i\nu}\, a^{(m_1,m_2|m|m_3,m_4)}_{\Delta_1,\Delta_2|-\tfrac{d}{2}+i\nu|\Delta_3,\Delta_4} \rho^{\text{AdS}}_{\Delta_1 \Delta_2 \Delta_3 \Delta_4}\left(\nu\right).
\end{multline}

\paragraph{The radial integral.} It is straightforward to evaluate the radial integral \eqref{aapp} in Mellin space. This is because in Mellin space the radial dependence is a power law and therefore reduces the radial integrals to Dirac delta functions \cite{Sleight:2020obc,Sleight:2021plv}. Inserting the Mellin representation \eqref{appgMB} for $g^{(m)}_{-\tfrac{d}{2}+i\nu}(R_1,R_2)$ and the radial kernel \eqref{MBK} this gives
\begin{multline}
  a^{(m_1,m_2|m|m_3,m_3)}_{\Delta_1,\Delta_2|-\tfrac{d}{2}+i\nu|\Delta_3,\Delta_4} = \frac{1}{2\pi}\int^{+i\infty}_{-i\infty}\frac{{\rm d}\omega}{2\pi i}\frac{{\rm d}{\bar \omega}}{2\pi i}\frac{{\rm d}\omega_1}{2\pi i}\frac{{\rm d}\omega_2}{2\pi i}\frac{{\rm d}\omega_3}{2\pi i}\frac{{\rm d}\omega_4}{2\pi i}\, \left(\frac{m}{2}\right)^{-2\left(\omega+{\bar \omega}\right)} \\ \times 2 \pi i\, \delta \left(d+2+\left(\tfrac{d}{2}+2\omega_1\right)+\left(\tfrac{d}{2}+2\omega_2\right)+\left(\tfrac{d}{2}+2\omega\right)\right) \\ \times 2 \pi i\, \delta \left(d+2+\left(\tfrac{d}{2}+2\omega_3\right)+\left(\tfrac{d}{2}+2\omega_4\right)+\left(\tfrac{d}{2}+2{\bar \omega}\right)\right) \\ \times \csc\left(\pi\left(\omega+{\bar \omega}\right) \right) \sin \left(\pi \left(\omega-\tfrac{i\nu}{2}\right)\right)\sin \left(\pi \left({\bar \omega}-\tfrac{i\nu}{2}\right)\right)\\ \times\Gamma\left(\omega+\tfrac{i\nu}{2}\right)\Gamma\left(\omega-\tfrac{i\nu}{2}\right)\Gamma\left(\bar{\omega}+\tfrac{i\nu}{2}\right)\Gamma\left(\bar{\omega}-\tfrac{i\nu}{2}\right)\\ \times \prod^4_{i=1}\frac{\Gamma\left(\omega_i+\tfrac{1}{2}\left(\Delta_i-\tfrac{d}{2}\right)\right)\Gamma\left(\omega_i-\tfrac{1}{2}\left(\Delta_i-\tfrac{d}{2}\right)\right)}{\Gamma\left(\tfrac{d}{2}-\Delta_i\right)}\left(\frac{m_i}{2}\right)^{-2\omega_i+\frac{d}{2}-\Delta_i}.
\end{multline}
Although it may look involved, the structural form of this expression is that of the Mellin-Barnes representation of a momentum space exchange Witten diagram in EAdS$_{1-d}$, where the masses $m_i$ and $m$ play the role of the modulus of the boundary momenta -- see \cite{Sleight:2020obc,Sleight:2021plv}. For massless external scalars (i.e. $m_i=0$ and generic $m$) this simplifies significantly and the Mellin integrals can be lifted:
\begin{multline}
  a^{(0,0|m|0,0)}_{\Delta_1,\Delta_2|-\tfrac{d}{2}+i\nu|\Delta_3,\Delta_4} = \frac{1}{2\pi}\int^{+i\infty}_{-i\infty}\frac{{\rm d}\omega}{2\pi i}\frac{{\rm d}{\bar \omega}}{2\pi i}\, \left(\frac{m}{2}\right)^{-2\left(\omega+{\bar \omega}\right)} \\ \times 2 \pi i\, \delta \left(\tfrac{d}{2}-2+2\omega-i\left(\nu_1+\nu_2\right)\right) 2 \pi i\, \delta \left(\tfrac{d}{2}-2+2{\bar \omega}-i\left(\nu_3+\nu_4\right)\right) \\ \times \csc\left(\pi\left(\omega+{\bar \omega}\right) \right) \sin \left(\pi \left(\omega-\tfrac{i\nu}{2}\right)\right)\sin \left(\pi \left({\bar \omega}-\tfrac{i\nu}{2}\right)\right)\\ \times\Gamma\left(\omega+\tfrac{i\nu}{2}\right)\Gamma\left(\omega-\tfrac{i\nu}{2}\right)\Gamma\left(\bar{\omega}+\tfrac{i\nu}{2}\right)\Gamma\left(\bar{\omega}-\tfrac{i\nu}{2}\right),
\end{multline}
where we used that for $m\to0$
\begin{multline}\label{mllim}
    \Gamma\left(\omega-\frac{1}{2}\left(\Delta - \tfrac{d}{2}\right)\right)\Gamma\left(\omega+\frac{1}{2}\left(\Delta - \tfrac{d}{2}\right)\right)\left(\frac{m}{2}\right)^{-2\omega+\frac{d}{2}-\Delta}\\ \to \Gamma\left(\tfrac{d}{2}-\Delta\right)\,2\pi i\, \delta \left(\omega+\frac{1}{2}\left(\Delta - \tfrac{d}{2}\right)\right).
\end{multline}
The remaining integrals in the Mellin variables $\omega$ and ${\bar \omega}$ can be evaluated using the Dirac delta functions, giving:
\begin{multline}
  a^{(0,0|m|0,0)}_{\Delta_1,\Delta_2|-\tfrac{d}{2}+i\nu|\Delta_3,\Delta_4} = \frac{1}{2\pi} \left(\frac{m}{2}\right)^{-4+3d-\sum\limits^4_{i=1}\Delta_i}  \csc\left(\frac{\pi}{2}\left(4-3d+\sum\limits^4_{i=1}\Delta_i\right) \right)\\ \times \sin \left(\tfrac{\pi}{2} \left(2-d+\Delta_1+\Delta_2-\left(\tfrac{d}{2}+i\nu\right)\right)\right)\sin \left(\tfrac{\pi}{2} \left(-\left(\tfrac{d}{2}+i\nu\right)+2-d+\Delta_3+\Delta_4\right)\right)\\ \times \Gamma \left(\tfrac{1}{2} \left(2-d+\Delta_1+\Delta_2-\left(\tfrac{d}{2}+i\nu\right)\right)\right)\Gamma \left(\tfrac{1}{2} \left(2-d+\Delta_3+\Delta_4-\left(\tfrac{d}{2}+i\nu\right)\right)\right)\\
  \times \Gamma \left(\tfrac{1}{2} \left(2-d+\Delta_1+\Delta_2-\left(\tfrac{d}{2}-i\nu\right)\right)\right)\Gamma \left(\tfrac{1}{2} \left(2-d+\Delta_3+\Delta_4-\left(\tfrac{d}{2}-i\nu\right)\right)\right).
\end{multline}

\section{Propagators for massless scalar fields}

In this appendix we give various results for propagators of massless scalar fields in Minkowski space.

\subsection{Celestial bulk-to-boundary propagator}
\label{A}

In this section we derive the expression \eqref{cbuboml2} for the celestial bulk-to-boundary propagator for massless scalars by applying the prescription \eqref{celestialbubo} directly to the massless Feynman propagator. In the main text the expression \eqref{cbuboml2} was obtained through the massless limit of the massive celestial bulk-to-boundary propagator and the results of this appendix confirms that the massless limit commutes with the prescription \eqref{celestialbubo}.

\vskip 4pt
Starting from the definition \eqref{celestialbubo}, the celestial bulk-to-boundary propagator for a massless scalar is defined as
\begin{equation}\label{celestialbuboapp}
G_\Delta^{(0)}(X,Q)=\lim_{\hat{Y}\to Q}\int_0^{\infty}\frac{{\rm d}R}{R}R^\Delta G_T^{(0)}(X,R\hat{Y}),
\end{equation}
where $G_T^{(0)}\left(X,Y\right)$ is the Feynman propagator for a massless scalar:
\begin{align}\label{appmlFeyn}
G_T^{(0)}(X,Y)&= \frac{\Gamma\left(\frac{d}{2}\right)}{4\pi^{\frac{d+2}{2}}}\frac{1}{\left[\left(X-Y\right)^2+i\epsilon\right]^{\frac{d}{2}}},\\
&=\frac{i^{-\frac{d}{2}}}{4\pi^{\frac{d+2}{2}}}\int^{\infty}_0 \frac{{\rm d}t}{t} t^{\frac{d}{2}} \exp\left[it\left(X-Y\right)^2\right],
\end{align}
where in the second equality we employed Schwinger-parameterisation. Inserting into the prescription \eqref{celestialbuboapp} we have (using that $Q^2=0$):
\begin{align}
G_\Delta^{(0)}(X,Q)&=\frac{i^{-\frac{d}{2}}}{4\pi^{\frac{d+2}{2}}}\int_0^{\infty}\frac{{\rm d}R}{R}R^\Delta \int_0^{\infty} \frac{{\rm d}t}{t} t^{\frac{d}{2}} \exp\left[itX^2-2itR X \cdot Q\right],\\
&=\frac{\Gamma\left(\Delta\right)}{4\pi^{\frac{d+2}{2}}}i^{\Delta-\frac{d}{2}}\left(-2X \cdot Q+i\epsilon\right)^{-\Delta} \int_0^{\infty} \frac{{\rm d}t}{t} t^{\frac{d}{2}} \exp\left[itX^2\right],\\
&=\frac{\Gamma\left(\Delta\right)\Gamma\left(\frac{d}{2}-\Delta\right)}{4\pi^{\frac{d+2}{2}}}\left(-2X \cdot Q+i\epsilon\right)^{-\Delta} \left(X^2+i\epsilon\right)^{\Delta-\frac{d}{2}},
\end{align}
which recovers the expression \eqref{cbuboml2} obtained via massless limit.

\subsection{Massless celestial two-point function} 
\label{C}

Likewise we can obtain the free celestial two-point function \eqref{c2ptml} starting from the massless Feynman propagator \eqref{appmlFeyn}. Applying the definition \eqref{bdry2pt} we have:
\begin{align}
G_\Delta^{(0)}(Q_1,Q_2)&=\lim_{\hat{X}\to Q_1}\lim_{\hat{Y}\to Q_2}\int_0^{\infty}\frac{{\rm d}R_2}{R_2}R_2^{\Delta_2}\int_0^{\infty}\frac{{\rm d}R_1}{R_1}R_1^{\Delta_1}G_T^{(0)}(R_1\hat{X},R_2\hat{Y}).
\end{align}
Using the Schwinger parameterisation and the Dirac delta function this gives
\begin{align}
G_\Delta^{(0)}(Q_1,Q_2)&=\frac{i^{-\frac{d}{2}}}{4\pi^{\frac{d+2}{2}}}\int_0^{\infty}\frac{{\rm d}R_2}{R_2}R_2^{\Delta_2}\int_0^{\infty}\frac{{\rm d}R_1}{R_1}R_1^{\Delta_1} \int_0^{\infty} \frac{{\rm d}t}{t} t^{\frac{d}{2}} \exp\left[-2itR_1 R_2 Q_1 \cdot Q_2\right],\\
&=\frac{\Gamma\left(\Delta_1\right)}{4\pi^{\frac{d+2}{2}}}i^{\Delta_1-\frac{d}{2}}\left(-2 Q_1 \cdot Q_2+i\epsilon\right)^{-\Delta_1}\int_0^{\infty}\frac{{\rm d}R_2}{R_2}R_2^{\Delta_2-\Delta_1} \int_0^{\infty} \frac{{\rm d}t}{t} t^{\frac{d}{2}-\Delta_1},\\
&=\frac{\Gamma\left(\Delta_1\right)}{4\pi^{\frac{d+2}{2}}}\left(-2 Q_1 \cdot Q_2+i\epsilon\right)^{-\Delta_1} \left(2\pi i\right)\delta\left(\Delta_1-\Delta_2\right)\left(2\pi i\right)\delta\left(\Delta_1-\tfrac{d}{2}\right),
\end{align}
which matches the equation \eqref{c2ptml} obtained by taking the massless limit of the massive celestial two-point function \eqref{bdry2pt}.

\section{Evaluating the spectral integral}
\label{app::spectralintegral}

In this appendix we show that the conformal partial wave expansion \eqref{cpwecexch} for the exchange of a massive scalar field between massless scalars recovers the expression given in \cite{Pacifico:2024dyo} using the Mellin representation of conformal correlators in position space upon evaluating the spectral integral.

\vskip 4pt 
The Mellin representation of conformal correlation functions was proposed by Mack in his seminal work \cite{Mack:2009mi,Mack:2009gy}. For a four-point function of operators $O_i$ takes the form: 
\begin{multline}\label{MBst}
\hspace*{-0.75cm}\left\langle O_{1}(Q_1)\ldots O_{4}(Q_4)\right\rangle = \frac{1}{\left(-2 Q_1 \cdot Q_2\right)^{\frac{\Delta_1+\Delta_2}{2}}\left(-2 Q_3 \cdot Q_4\right)^{\frac{\Delta_3+\Delta_4}{2}}}\left(\frac{Q_2 \cdot Q_4}{Q_1 \cdot Q_4}\right)^{\frac{\Delta_1-\Delta_2}{2}}\left(\frac{Q_1 \cdot Q_4}{Q_1 \cdot Q_3}\right)^{\frac{\Delta_3-\Delta_4}{2}}\\ \times \int^{+i\infty}_{-i\infty}\frac{{\rm d}s{\rm d}t}{\left(4\pi i\right)^2}\, u^{\frac{t}{2}}v^{-\left(\frac{s+t}{2}\right)}\,M\left(s,t\right)\Gamma\left(\frac{\Delta_1+\Delta_2-t}{2}\right)\Gamma\left(\frac{\Delta_3+\Delta_4-t}{2}\right)\\ \times \Gamma\left(\frac{\Delta_{34}-s}{2}\right)\Gamma\left(\frac{-\Delta_{12}-s}{2}\right)\Gamma\left(\frac{s+t}{2}\right)\Gamma\left(\frac{s+t+\Delta_{12}-\Delta_{34}}{2}\right),
\end{multline}
where $\Delta_{ij}=\Delta_i-\Delta_j$ and conformal invariant cross ratios
\begin{equation}
    u = \frac{Q_1 \cdot Q_2\,Q_3 \cdot Q_4}{Q_1 \cdot Q_3 Q_2 \cdot Q_4}, \qquad v = \frac{Q_1 \cdot Q_4\,Q_2 \cdot Q_3}{Q_1 \cdot Q_3 Q_2 \cdot Q_4}.
\end{equation}
The function $M\left(s,t\right)$ is the corresponding Mellin amplitude which is a function of the Mellin variables $s$ and $t$. The work \cite{Pacifico:2024dyo} proposed to use the Mellin representation \eqref{MBst} to study celestial correlators \eqref{ccdefn}, where the corresponding Mellin amplitudes where dubbed \emph{celestial Mellin amplitudes}.

\vskip 4pt
Starting from the conformal partial wave expansion \eqref{cpwecexch} for the exchange diagram, the corresponding Mellin amplitude reads
\begin{equation}
   M^{(0,0|m|0,0)}_{\Delta_1 \Delta_2 \Delta_3 \Delta_4}\left(s,t\right) = \int^{\infty}_{-\infty}\frac{{\rm d}\nu}{2\pi}\,\rho^{(0,0|m|0,0)}_{\Delta_1 \Delta_2 \Delta_3 \Delta_4}\left(\nu\right){\cal F}^{12,34}_{\nu,0}\left(s,t\right),
\end{equation}
where the Mellin representation of the conformal partial wave is: 
\begin{multline}
    {\cal F}^{12,34}_{\nu,0}\left(s,t\right) = \tfrac{\pi ^{d/2}}{2 \Gamma \left(\frac{d+2 \Delta_1-2 \Delta_2+2 i \nu}{4}\right) \Gamma \left(\frac{d-2 \Delta_1+2 \Delta_2+2 i \nu}{4}\right) \Gamma \left(\frac{d+2 \Delta_3-2 \Delta_4-2 i \nu}{4} \right) \Gamma \left(\frac{d-2 \Delta_3+2 \Delta_4-2 i \nu }{4}\right)}\\\times\frac{\Gamma \left(\frac{d-2 (t+i \nu )}{4}\right) \Gamma \left(\frac{d-2 t+2 i \nu }{4}\right)}{\Gamma \left(\frac{-t+\Delta_1+\Delta_2}{2}\right) \Gamma \left(\frac{-t+\Delta_3+\Delta_4}{2}\right)}.
\end{multline}

At the level of the Mellin representation the spectral integral takes the form:
\begin{multline}\label{generalSI}
  \hspace*{-0.5cm}  I_{a_1,a_2,a_3,a_4,a_5}=\int_{-\infty}^{\infty}\frac{{\rm d}\nu}{2\pi}\,\left[\prod_{j=1}^5\Gamma\left(a_j+\tfrac{i\nu}2\right)\Gamma\left(a_j-\tfrac{i\nu}2\right)\right] \left[-2i\nu\prod_{j=2}^5\frac1{\pi}\sin\left(\pi(a_j-\tfrac{i\nu}2\right)\right]\,,
\end{multline}
with 
\begin{subequations}\label{exchai}
\begin{align}
    a_1&=\frac{d-2t}{4}\,,\\
    a_2&=\frac{2\Delta_{1}+2\Delta_{2}-d}{4}\,,\\
    a_3&=\frac{2\Delta_{3}+2\Delta_{4}-d}{4}\,,\\
    a_4&=\frac{2\Delta_{1}+2\Delta_{2}-3d+4}{4}\,,\\
    a_5&=\frac{2\Delta_{3}+2\Delta_{4}-3d+4}{4}\,.
\end{align}
\end{subequations}
The above choice of parameters allows us to see that $a_i$ are not all independent and one can for instance set $a_5=-a_2+a_3+a_4$, but to keep the discussion general we shall not do so in the following. The integral also is manifestly symmetric under permutation of $a_j$, $j=2,..,4$ and we can take advantage of this to permute the latter coefficients and obtain different expressions for the same result.

\vskip 4pt
In order to evaluate the spectral integral \eqref{generalSI}, writing $\nu=R e^{i\theta}$ we first note the asymptotic behaviour of the integrand as $R\to\infty$ is given by:
\begin{align}
e^{-\frac{1}{2} \pi  R | \cos (\theta )| } R^{2 \text{Re}(a_1+a_2+a_3+a_4+a_5)-4}\,.
\end{align}
This makes the integral exponentially suppressed apart for the imaginary axis where it becomes power-law. Nonetheless one can identify a convergence region for the integral by the condition:
\begin{align}
    2\text{Re}(a_1+a_2+a_3+a_4+a_5)<3.
\end{align}
Taking $a_i$ to be the values \eqref{exchai} as they appear in the exchange diagram, for principal series representations $\Delta_i = \frac{d}{2}+i\nu_i$, $\nu \in \mathbb{R}$, we have  
\begin{equation}
   2\text{Re}(a_1+a_2+a_3+a_4+a_5)=2+\frac{d}{4}-\frac{1}{2}\text{Re}(t). 
\end{equation}
This bound is saturated for $d=2$, in which case one needs to pick an appropriate contour for $t$ (or equivalently an $i\epsilon$ prescription). This issue will however not play an important role since the Mellin representation easily allows for such a choice of contour.\footnote{Alternatively one could define the spectral integral with a slight tilt at $\infty$ in order to maintain the integral convergent.}

\vskip 4pt
Having discussed the general properties of the spectral integral, in the following we evaluate it by applying Cauchy's residue theorem. Closing the contour in the lower-half plane encloses a single series of poles at:
\begin{equation}
    \nu = -2i(a_1+n), \qquad n=0, 1, 2, \ldots\,.
\end{equation}
This gives the result as a difference of ${}_5F_4$ hypergeometric functions evaluated at argument $z=-1$:\footnote{The tilde denotes the regularised hypergeometric function:
\begin{equation}
    {}_p{\tilde F}_q\left(a_1,\ldots,a_p;b_1, \ldots, b_q;z\right)=\frac{ {}_pF_q\left(a_1,\ldots,a_p;b_1, \ldots, b_q;z\right)}{\Gamma\left(b_1\right) \ldots \Gamma\left(b_q\right)}.
\end{equation}
}
\begin{align}
    I_{a_1,a_2,a_3,a_4,a_5}&=-4 \Gamma (2 a_1+1) \Gamma (a_1+a_2) \Gamma (a_1+a_3) \Gamma (a_1+a_4) \Gamma (a_1+a_5)\\
    \Bigg[
&\, _5\tilde{F}_4\left(\begin{matrix}
    2 a_1,a_1+a_2,a_1+a_3,a_1+a_4,a_1+a_5\\\nonumber
    a_1-a_2+1,a_1-a_3+1,a_1-a_4+1,a_1-a_5+1
\end{matrix};-1\right)\\
-&2 (a_1+a_2) (a_1+a_3) (a_1+a_4) (a_1+a_5) \,\\\nonumber
\times&_5\tilde{F}_4\left(\begin{matrix}
    2 a_1+1,a_1+a_2+1,a_1+a_3+1,a_1+a_4+1,a_1+a_5+1\\\nonumber
    a_1-a_2+2,a_1-a_3+2,a_1-a_4+2,a_1-a_5+2
\end{matrix};-1\right)
    \Bigg].
\end{align}
The above expression is not easy to use: Naively it has poles at $a_1=-\frac{n+1}{2}$, $n \in \mathbb{N}$, which would correspond to poles at Mellin variable at $t=\frac{d}{2}-n$. These however cancel in a non-trivial way in the difference of hypergeometric functions. A simpler form however can be obtained using standard Mellin-Barnes methods to derive transformation formulas for hypergeometric functions. To this end, consider Barnes second lemma:
\begin{multline}
    \int^{i\infty}_{-i\infty}\frac{{\rm d}s}{2\pi i}\,\frac{\Gamma (-s) \Gamma (a+s) \Gamma (b+s) \Gamma (c+s) \Gamma (-d-s+1)}{\Gamma (e+s)}\\=\frac{\Gamma (a) \Gamma (b) \Gamma (c) \Gamma (a-d+1) \Gamma (b-d+1) \Gamma (c-d+1)}{\Gamma (e-a) \Gamma (e-b) \Gamma (e-c)}\,,
\end{multline}
where $e=a + b + c - d + 1$. Picking
\begin{align}
    a&=a_1+\frac{i \nu}2\,,\\
    b&=a_4+\frac{i \nu}2\,,\\
    c&=a_5+\frac{i \nu}2\,,\\
    d&=1+i\nu\,,
\end{align}
this is
\begin{multline}
    \int^{i\infty}_{-i\infty}\frac{{\rm d}s}{2\pi i}\,\frac{\Gamma (a_1+s) \Gamma (a_4+s) \Gamma (a_5+s) \Gamma \left(-s-\frac{i \nu }{2}\right) \Gamma \left(-s+\frac{i \nu }{2}\right)}{\Gamma (a_1+a_4+a_5+s)}\\=\frac{\Gamma \left(a_1-\frac{i \nu }{2}\right) \Gamma \left(a_1+\frac{i \nu }{2}\right) \Gamma \left(a_4-\frac{i \nu }{2}\right) \Gamma \left(a_4+\frac{i \nu }{2}\right) \Gamma \left(a_5-\frac{i \nu }{2}\right) \Gamma \left(a_5+\frac{i \nu }{2}\right)}{\Gamma (a_1+a_4) \Gamma (a_1+a_5) \Gamma (a_4+a_5)}\,.
\end{multline}
Using the above identities one can trade six of the ten $\Gamma$-functions in the spectral integral \eqref{generalSI} for just two. This gives the spectral integral in terms of more manageable ${}_3F_2$ hypergeometric functions at argument $z=-1$:
\begin{align}
    I_{a_1,a_2,a_3,a_4,a_5}&= \int^{i\infty}_{-i\infty}\frac{{\rm d}s}{2\pi i}\,\frac{4 \Gamma (1-2 s) \Gamma (a_1+a_4) \Gamma (a_1+a_5) \Gamma (a_1+s) \Gamma (a_2-s) \Gamma (a_3-s) \Gamma (a_4+a_5)}{\Gamma (-a_4-s+1) \Gamma (-a_5-s+1) \Gamma (a_1+a_4+a_5+s)}\nonumber\\\nonumber
    &\times \Bigg[2 (a_2-s) (a_3-s) \, _3\tilde{F}_2\left(\begin{matrix}
        1-2 s,a_2-s+1,a_3-s+1\\
        -a_2-s+2,-a_3-s+2
    \end{matrix};-1\right)\\
    &\hspace{130pt}-\, _3\tilde{F}_2\left(\begin{matrix}
        -2s,a_2-s,a_3-s\\
        -a_2-s+1,-a_3-s+1
    \end{matrix};-1\right)\Bigg]\,.
\end{align}
The remaining Mellin integral can be evaluated exactly by closing the contour on the right on the poles:
\begin{align}
    s &= a_2+n, \qquad n = 0, 1, 2, \ldots\,, \\
    s &= a_3+n, \qquad n = 0, 1, 2, \ldots\,,
\end{align}
which gives
\begin{align}
    I_{a_1,a_2,a_3,a_4,a_5}&=-\frac{4\Gamma (a_1+a_4)\Gamma (a_1+a_5) \Gamma (a_4+a_5)}{\pi^2\,\Gamma (1-a_2-a_3)}\left(\sum_{n=0}^\infty R_{a_2+n}+\sum_{n=0}^\infty R_{a_3+n}\right)\,,
\end{align}
where
\begin{align}
    R_{a_2+n}&=\sin (\pi  (a_2+a_4)) \sin (\pi  (a_2+a_5)) \\\nonumber
    &\hspace{50pt}\times\frac{(-1)^n \Gamma (a_1+a_2+n) \Gamma (-a_2+a_3-n) \Gamma (a_2+a_4+n) \Gamma (a_2+a_5+n))}{n!  \Gamma (a_1+a_2+a_4+a_5+n)}\,,\\\nonumber
    R_{a_3+n}&=\sin (\pi  (a_3+a_4) \sin (\pi  (a_3+a_5)) \\
    &\hspace{50pt}\times\frac{(-1)^n\Gamma (a_1+a_3+n) \Gamma (a_2-a_3-n) \Gamma (a_3+a_4+n) \Gamma (a_3+a_5+n))}{n!  \Gamma (a_1+a_3+a_4+a_5+n)}\,.
\end{align}
Carrying out the summation and exchanging without loss of generality $a_2\leftrightarrow a_4$ gives the following expression in terms of ${}_3F_2$ evaluated at $z=1$:
\begin{align}\label{ResFin}
    I_{a_1,a_2,a_3,a_4,a_5}&=\frac{4 \pi \csc (\pi  (a_3-a_4)) \Gamma (a_1+a_2) \Gamma (a_1+a_5) \Gamma (a_2+a_5)}{\Gamma (1-a_3-a_4)}\\\nonumber
    &\hspace{50pt}\times\Bigg[\frac{\Gamma (a_1+a_3)}{\Gamma (1-a_2-a_3) \Gamma (1-a_3-a_5)}\\\nonumber
    &\hspace{120pt}\times \, _3\tilde{F}_2\left(\begin{matrix}
        a_1+a_3,a_2+a_3,a_3+a_5\\
        1+a_3-a_4,a_1+a_2+a_3+a_5
    \end{matrix};1\right)\\\nonumber
    &\hspace{50pt}-\frac{\Gamma (a_1+a_4)}{\Gamma (1-a_2-a_4) \Gamma (1-a_4-a_5)}\\\nonumber
    &\hspace{120pt}\times \,
    \, _3\tilde{F}_2\left(\begin{matrix}
        a_1+a_4,a_2+a_4,a_4+a_5\\
        1-a_3+a_4,a_1+a_2+a_4+a_5
    \end{matrix};1\right)
    \Bigg]\,.
\end{align}
Finally, we claim that this is equivalent to the much simpler expression:
\begin{shaded}
\begin{align}\label{FinalRes}
    I_{a_1,a_2,a_3,a_4,a_5}&=-\frac{4 \Gamma (a_1+a_2) \Gamma (a_1+a_3) \Gamma (a_1+a_4) \Gamma (a_1+a_5)}{\Gamma (1-a_4-a_5)}\\\nonumber
    &\hspace{150pt}\times \, _3\tilde{F}_2\left(\begin{matrix}
        a_1+a_4,a_1+a_5,1-a_2-a_3\\
        a_1-a_2+1,a_1-a_3+1
    \end{matrix};1\right)\,.
\end{align}
\end{shaded}
This is obtained using the following standard transformation formulas:
\begin{subequations}
\begin{align}\label{Rule3F2a}
    _3F_2&\left(\begin{matrix}
        a_1,a_2,a_3\\
        b_1,b_2
    \end{matrix};1\right)=\frac{\Gamma (b_1) \Gamma (-a_1-a_2+b_1)}{\Gamma (b_1-a_1) \Gamma (b_1-a_2)}\, _3F_2\left(\begin{matrix}
        a_1,a_2,b_2-a_3\\
        a_1+a_2-b_1+1,b_2
    \end{matrix};1\right)\\\nonumber
    &\hspace{50pt}+\frac{\Gamma (b_1) \Gamma (b_2) \Gamma (a_1+a_2-b_1) \Gamma (-a_1-a_2-a_3+b_1+b_2)}{\Gamma (a_1) \Gamma (a_2) \Gamma (b_2-a_3) \Gamma (-a_1-a_2+b_1+b_2)}\\\nonumber
    &\hspace{100pt}\times\,_3F_2\left(\begin{matrix}
        b_1-a_1,b_1-a_2,-a_1-a_2-a_3+b_1+b_2\\
        -a_1-a_2+b_1+1,-a_1-a_2+b_1+b_2
    \end{matrix};1\right)\,,\\
\label{Gauss3F2a}
    _3F_2&\left(\begin{matrix}
        a_1,a_2,a_3\\
        b_1,b_2
    \end{matrix};1\right)=\frac{\Gamma (b_1) \Gamma (-a_1-a_2-a_3+b_1+b_2) }{\Gamma (b_1-a_1) \Gamma (-a_2-a_3+b_1+b_2)} \\\nonumber
    &\hspace{150pt}\times\, _3F_2\left(\begin{matrix}
        a_1,b_2-a_2,b_2-a_3\\
        b_2,-a_2-a_3+b_1+b_2
    \end{matrix};1\right)\,,\\
\label{Gauss3F2b}
    _3F_2&\left(\begin{matrix}
        a_1,a_2,a_3\\
        b_1,b_2
    \end{matrix};1\right)=\frac{\Gamma (b_1) \Gamma (b_2) \Gamma (-a_1-a_2-a_3+b_1+b_2)}{\Gamma (a_1) \Gamma (-a_1-a_2+b_1+b_2) \Gamma (-a_1-a_3+b_1+b_2)}\\\nonumber
    &\hspace{90pt}\times\, _3F_2\left(\begin{matrix}
        b_1-a_1,b_2-a_1,-a_1-a_2-a_3+b_1+b_2\\
        -a_1-a_2+b_1+b_2,-a_1-a_3+b_1+b_2
    \end{matrix};1\right).
\end{align}
\end{subequations}
Applying \eqref{Rule3F2a} followed by \eqref{Gauss3F2a} on \eqref{FinalRes} we can see that we get a sum of two ${}_3F_2$ of which one summand exactly matches the first summand in \eqref{ResFin}. One is then left to show that also the other summands is equal to the second term in \eqref{ResFin}. This can be done applying \eqref{Gauss3F2b} followed by \eqref{Gauss3F2a} on the second term in \eqref{ResFin}, concluding the proof.

\vskip 4pt
The expression \eqref{FinalRes} with \eqref{exchai} recovers the Mellin amplitude for the exchange diagram given in \cite{Pacifico:2024dyo}:
\begin{multline}
     M^{(0,0|m|0,0)}_{\Delta_1 \Delta_2 \Delta_3 \Delta_4}\left(s,t\right)=g_{12}g_{34}\,i^{-\frac{1}{2}\sum_i \Delta_i}\pi^{d+2}\prod^4_{i=1}\frac{1}{4 \pi^{\frac{d+2}{2}}}\Gamma\left(\tfrac{d}{2}-\Delta_i\right)\,\\ \times \frac{1}{4\pi^{\frac{d+2}{2}}} \left(\frac{m}{2}\right)^{3d-4-\sum_i\Delta_i}\Gamma\left(\tfrac{-3d+4+\sum_i\Delta_i}{2}\right) \frac{\Gamma\left(\tfrac{2-d+\Delta_1+\Delta_2-s_{12}}{2}\right)\Gamma\left(\tfrac{2-d+\Delta_3+\Delta_4-s_{12}}{2}\right)}{\Gamma\left(\tfrac{d+2-\Delta_1-\Delta_2-s_{12}}{2}\right)\Gamma\left(\tfrac{d+2-\Delta_3-\Delta_4-s_{12}}{2}\right)}\,\\ \times {}_3F_2\left(\begin{matrix}\frac{d+2- \Delta_1-\Delta_2-\Delta_3-\Delta_4}{2},\tfrac{2-d+\Delta_1+\Delta_2-t}{2},\tfrac{2-d+\Delta_3+\Delta_4-t}{2}\\\tfrac{d+2-\Delta_3-\Delta_4-t}{2},\tfrac{d+2-\Delta_1-\Delta_2-t}{2}\end{matrix};1\right).
\end{multline}

\end{appendix}

\bibliographystyle{JHEP}
\bibliography{refs}

\providecommand{\href}[2]{#2}\begingroup\raggedright\begin{thebibliography}{10}

\bibitem{Sleight:2023ojm}
C.~Sleight and M.~Taronna, \emph{{Celestial Holography Revisited}}, \href{https://doi.org/10.1103/PhysRevLett.133.241601}{\emph{Phys. Rev. Lett.} {\bfseries 133} (2024) 241601} [\href{https://arxiv.org/abs/2301.01810}{{\ttfamily 2301.01810}}].

\bibitem{Raclariu:2021zjz}
A.-M. Raclariu, \emph{{Lectures on Celestial Holography}},  \href{https://arxiv.org/abs/2107.02075}{{\ttfamily 2107.02075}}.

\bibitem{Pasterski:2021rjz}
S.~Pasterski, \emph{{Lectures on celestial amplitudes}}, \href{https://doi.org/10.1140/epjc/s10052-021-09846-7}{\emph{Eur. Phys. J. C} {\bfseries 81} (2021) 1062} [\href{https://arxiv.org/abs/2108.04801}{{\ttfamily 2108.04801}}].

\bibitem{McLoughlin:2022ljp}
T.~McLoughlin, A.~Puhm and A.-M. Raclariu, \emph{{The SAGEX Review on Scattering Amplitudes, Chapter 11: Soft Theorems and Celestial Amplitudes}},  \href{https://arxiv.org/abs/2203.13022}{{\ttfamily 2203.13022}}.

\bibitem{Pasterski:2021raf}
S.~Pasterski, M.~Pate and A.-M. Raclariu, \emph{{Celestial Holography}},  in \emph{{2022 Snowmass Summer Study}}, 11, 2021, \href{https://arxiv.org/abs/2111.11392}{{\ttfamily 2111.11392}}.

\bibitem{Pasterski:2016qvg}
S.~Pasterski, S.-H. Shao and A.~Strominger, \emph{{Flat Space Amplitudes and Conformal Symmetry of the Celestial Sphere}}, \href{https://doi.org/10.1103/PhysRevD.96.065026}{\emph{Phys. Rev.} {\bfseries D96} (2017) 065026} [\href{https://arxiv.org/abs/1701.00049}{{\ttfamily 1701.00049}}].

\bibitem{Pasterski:2017kqt}
S.~Pasterski and S.-H. Shao, \emph{{Conformal basis for flat space amplitudes}}, \href{https://doi.org/10.1103/PhysRevD.96.065022}{\emph{Phys. Rev.} {\bfseries D96} (2017) 065022} [\href{https://arxiv.org/abs/1705.01027}{{\ttfamily 1705.01027}}].

\bibitem{Maldacena:1997re}
J.~M. Maldacena, \emph{{The Large N limit of superconformal field theories and supergravity}}, \href{https://doi.org/10.1023/A:1026654312961, 10.4310/ATMP.1998.v2.n2.a1}{\emph{Int. J. Theor. Phys.} {\bfseries 38} (1999) 1113} [\href{https://arxiv.org/abs/hep-th/9711200}{{\ttfamily hep-th/9711200}}].

\bibitem{Gubser:1998bc}
S.~S. Gubser, I.~R. Klebanov and A.~M. Polyakov, \emph{{Gauge theory correlators from noncritical string theory}}, \href{https://doi.org/10.1016/S0370-2693(98)00377-3}{\emph{Phys. Lett.} {\bfseries B428} (1998) 105} [\href{https://arxiv.org/abs/hep-th/9802109}{{\ttfamily hep-th/9802109}}].

\bibitem{Witten:1998qj}
E.~Witten, \emph{{Anti-de Sitter space and holography}}, \href{https://doi.org/10.4310/ATMP.1998.v2.n2.a2}{\emph{Adv. Theor. Math. Phys.} {\bfseries 2} (1998) 253} [\href{https://arxiv.org/abs/hep-th/9802150}{{\ttfamily hep-th/9802150}}].

\bibitem{deBoer:2003vf}
J.~de~Boer and S.~N. Solodukhin, \emph{{A Holographic reduction of Minkowski space-time}}, \href{https://doi.org/10.1016/S0550-3213(03)00494-2}{\emph{Nucl. Phys. B} {\bfseries 665} (2003) 545} [\href{https://arxiv.org/abs/hep-th/0303006}{{\ttfamily hep-th/0303006}}].

\bibitem{Cheung:2016iub}
C.~Cheung, A.~de~la Fuente and R.~Sundrum, \emph{{4D scattering amplitudes and asymptotic symmetries from 2D CFT}}, \href{https://doi.org/10.1007/JHEP01(2017)112}{\emph{JHEP} {\bfseries 01} (2017) 112} [\href{https://arxiv.org/abs/1609.00732}{{\ttfamily 1609.00732}}].

\bibitem{Casali:2022fro}
E.~Casali, W.~Melton and A.~Strominger, \emph{{Celestial amplitudes as AdS-Witten diagrams}}, \href{https://doi.org/10.1007/JHEP11(2022)140}{\emph{JHEP} {\bfseries 11} (2022) 140} [\href{https://arxiv.org/abs/2204.10249}{{\ttfamily 2204.10249}}].

\bibitem{Iacobacci:2022yjo}
L.~Iacobacci, C.~Sleight and M.~Taronna, \emph{{From celestial correlators to AdS, and back}}, \href{https://doi.org/10.1007/JHEP06(2023)053}{\emph{JHEP} {\bfseries 06} (2023) 053} [\href{https://arxiv.org/abs/2208.01629}{{\ttfamily 2208.01629}}].

\bibitem{Melton:2023bjw}
W.~Melton, A.~Sharma and A.~Strominger, \emph{{Celestial leaf amplitudes}}, \href{https://doi.org/10.1007/JHEP07(2024)132}{\emph{JHEP} {\bfseries 07} (2024) 132} [\href{https://arxiv.org/abs/2312.07820}{{\ttfamily 2312.07820}}].

\bibitem{Melton:2024jyq}
W.~Melton, A.~Sharma and A.~Strominger, \emph{{Soft algebras for leaf amplitudes}}, \href{https://doi.org/10.1007/JHEP07(2024)070}{\emph{JHEP} {\bfseries 07} (2024) 070} [\href{https://arxiv.org/abs/2402.04150}{{\ttfamily 2402.04150}}].

\bibitem{deGioia:2022fcn}
L.~P. de~Gioia and A.-M. Raclariu, \emph{{Eikonal approximation in celestial CFT}}, \href{https://doi.org/10.1007/JHEP03(2023)030}{\emph{JHEP} {\bfseries 03} (2023) 030} [\href{https://arxiv.org/abs/2206.10547}{{\ttfamily 2206.10547}}].

\bibitem{deGioia:2023cbd}
L.~P. de~Gioia and A.-M. Raclariu, \emph{{Celestial Sector in CFT: Conformally Soft Symmetries}}, \href{https://doi.org/10.21468/SciPostPhys.17.1.002}{\emph{SciPost Phys.} {\bfseries 17} (2024) 002} [\href{https://arxiv.org/abs/2303.10037}{{\ttfamily 2303.10037}}].

\bibitem{Iacobacci:2024nhw}
L.~Iacobacci, C.~Sleight and M.~Taronna, \emph{{Celestial holography revisited. Part II. Correlators and K\"all\'en-Lehmann}}, \href{https://doi.org/10.1007/JHEP08(2024)033}{\emph{JHEP} {\bfseries 08} (2024) 033} [\href{https://arxiv.org/abs/2401.16591}{{\ttfamily 2401.16591}}].

\bibitem{Dobrev:1975ru}
V.~K. Dobrev, V.~B. Petkova, S.~G. Petrova and I.~T. Todorov, \emph{{Dynamical Derivation of Vacuum Operator Product Expansion in Euclidean Conformal Quantum Field Theory}}, \href{https://doi.org/10.1103/PhysRevD.13.887}{\emph{Phys. Rev.} {\bfseries D13} (1976) 887}.

\bibitem{Dobrev:1977qv}
V.~Dobrev, G.~Mack, V.~Petkova, S.~Petrova and I.~Todorov, \emph{{Harmonic Analysis on the n-Dimensional Lorentz Group and Its Application to Conformal Quantum Field Theory}}, vol.~63. 1977, \href{https://doi.org/10.1007/BFb0009678}{10.1007/BFb0009678}.

\bibitem{Mack:2009mi}
G.~Mack, \emph{{D-independent representation of Conformal Field Theories in D dimensions via transformation to auxiliary Dual Resonance Models. Scalar amplitudes}},  \href{https://arxiv.org/abs/0907.2407}{{\ttfamily 0907.2407}}.

\bibitem{Costa:2012cb}
M.~S. Costa, V.~Gon\c{c}alves and J.~Penedones, \emph{{Conformal Regge theory}}, \href{https://doi.org/10.1007/JHEP12(2012)091}{\emph{JHEP} {\bfseries 12} (2012) 091} [\href{https://arxiv.org/abs/1209.4355}{{\ttfamily 1209.4355}}].

\bibitem{Caron-Huot:2017vep}
S.~Caron-Huot, \emph{{Analyticity in Spin in Conformal Theories}}, \href{https://doi.org/10.1007/JHEP09(2017)078}{\emph{JHEP} {\bfseries 09} (2017) 078} [\href{https://arxiv.org/abs/1703.00278}{{\ttfamily 1703.00278}}].

\bibitem{Lam:2017ofc}
H.~T. Lam and S.-H. Shao, \emph{{Conformal Basis, Optical Theorem, and the Bulk Point Singularity}}, \href{https://doi.org/10.1103/PhysRevD.98.025020}{\emph{Phys. Rev. D} {\bfseries 98} (2018) 025020} [\href{https://arxiv.org/abs/1711.06138}{{\ttfamily 1711.06138}}].

\bibitem{Nandan:2019jas}
D.~Nandan, A.~Schreiber, A.~Volovich and M.~Zlotnikov, \emph{{Celestial Amplitudes: Conformal Partial Waves and Soft Limits}}, \href{https://doi.org/10.1007/JHEP10(2019)018}{\emph{JHEP} {\bfseries 10} (2019) 018} [\href{https://arxiv.org/abs/1904.10940}{{\ttfamily 1904.10940}}].

\bibitem{Atanasov:2021cje}
A.~Atanasov, W.~Melton, A.-M. Raclariu and A.~Strominger, \emph{{Conformal block expansion in celestial CFT}}, \href{https://doi.org/10.1103/PhysRevD.104.126033}{\emph{Phys. Rev. D} {\bfseries 104} (2021) 126033} [\href{https://arxiv.org/abs/2104.13432}{{\ttfamily 2104.13432}}].

\bibitem{Melton:2021kkz}
W.~Melton, \emph{{Celestial Feynman Rules for Scalars}},  \href{https://arxiv.org/abs/2109.07462}{{\ttfamily 2109.07462}}.

\bibitem{Chang:2023ttm}
C.-M. Chang, R.~Liu and W.-J. Ma, \emph{{Split representation in celestial holography}},  \href{https://arxiv.org/abs/2311.08736}{{\ttfamily 2311.08736}}.

\bibitem{Fan:2023lky}
W.~Fan, \emph{{Celestial conformal blocks of massless scalars and analytic continuation of the Appell function F$_{1}$}}, \href{https://doi.org/10.1007/JHEP01(2024)145}{\emph{JHEP} {\bfseries 01} (2024) 145} [\href{https://arxiv.org/abs/2311.11345}{{\ttfamily 2311.11345}}].

\bibitem{Liu:2024vmx}
R.~Liu and W.-J. Ma, \emph{{Celestial Optical Theorem}},  \href{https://arxiv.org/abs/2404.18898}{{\ttfamily 2404.18898}}.

\bibitem{Kulp:2024scx}
J.~Kulp and S.~Pasterski, \emph{{Multiparticle States for the Flat Hologram}},  \href{https://arxiv.org/abs/2501.00462}{{\ttfamily 2501.00462}}.

\bibitem{Moschella:2007zza}
U.~Moschella and R.~Schaeffer, \emph{{Quantum theory on Lobatchevski spaces}}, \href{https://doi.org/10.1088/0264-9381/24/14/003}{\emph{Class. Quant. Grav.} {\bfseries 24} (2007) 3571} [\href{https://arxiv.org/abs/0709.2795}{{\ttfamily 0709.2795}}].

\bibitem{Penedones:2007ns}
J.~Penedones, \emph{{High Energy Scattering in the AdS/CFT Correspondence}}, Ph.D. thesis, Porto U., 2007.
\newblock \href{https://arxiv.org/abs/0712.0802}{{\ttfamily 0712.0802}}.

\bibitem{Costa:2014kfa}
M.~S. Costa, V.~Gon\c{c}alves and J.~Penedones, \emph{{Spinning AdS Propagators}}, \href{https://doi.org/10.1007/JHEP09(2014)064}{\emph{JHEP} {\bfseries 09} (2014) 064} [\href{https://arxiv.org/abs/1404.5625}{{\ttfamily 1404.5625}}].

\bibitem{Penedones:2010ue}
J.~Penedones, \emph{{Writing CFT correlation functions as AdS scattering amplitudes}}, \href{https://doi.org/10.1007/JHEP03(2011)025}{\emph{JHEP} {\bfseries 03} (2011) 025} [\href{https://arxiv.org/abs/1011.1485}{{\ttfamily 1011.1485}}].

\bibitem{ElShowk:2011ag}
S.~El-Showk and K.~Papadodimas, \emph{{Emergent Spacetime and Holographic CFTs}}, \href{https://doi.org/10.1007/JHEP10(2012)106}{\emph{JHEP} {\bfseries 10} (2012) 106} [\href{https://arxiv.org/abs/1101.4163}{{\ttfamily 1101.4163}}].

\bibitem{Pacifico:2024dyo}
F.~Pacifico, P.~Pergola and C.~Sleight, \emph{{Celestial Mellin Amplitudes}},  \href{https://arxiv.org/abs/2412.11992}{{\ttfamily 2412.11992}}.

\bibitem{Sleight:2019mgd}
C.~Sleight, \emph{{A Mellin Space Approach to Cosmological Correlators}}, {\emph{JHEP} {\bfseries 01} (2020) 090} [\href{https://arxiv.org/abs/1906.12302}{{\ttfamily 1906.12302}}].

\bibitem{Sleight:2019hfp}
C.~Sleight and M.~Taronna, \emph{{Bootstrapping Inflationary Correlators in Mellin Space}}, \href{https://doi.org/10.1007/JHEP02(2020)098}{\emph{JHEP} {\bfseries 02} (2020) 098} [\href{https://arxiv.org/abs/1907.01143}{{\ttfamily 1907.01143}}].

\bibitem{Sleight:2020obc}
C.~Sleight and M.~Taronna, \emph{{From AdS to dS exchanges: Spectral representation, Mellin amplitudes, and crossing}}, \href{https://doi.org/10.1103/PhysRevD.104.L081902}{\emph{Phys. Rev. D} {\bfseries 104} (2021) L081902} [\href{https://arxiv.org/abs/2007.09993}{{\ttfamily 2007.09993}}].

\bibitem{Sleight:2021plv}
C.~Sleight and M.~Taronna, \emph{{From dS to AdS and back}}, \href{https://doi.org/10.1007/JHEP12(2021)074}{\emph{JHEP} {\bfseries 12} (2021) 074} [\href{https://arxiv.org/abs/2109.02725}{{\ttfamily 2109.02725}}].

\bibitem{Fronsdal:1978vb}
C.~Fronsdal, \emph{{Singletons and Massless, Integral Spin Fields on de Sitter Space (Elementary Particles in a Curved Space. 7.}}, \href{https://doi.org/10.1103/PhysRevD.20.848}{\emph{Phys. Rev.} {\bfseries D20} (1979) 848}.

\bibitem{toappear}
S.~Malherbe, P.~Pergola, C.~Sleight and M.~Taronna, ``{Spinning Celestial Mellin Amplitudes}.'' To appear.

\bibitem{Freedman:1998tz}
D.~Z. Freedman, S.~D. Mathur, A.~Matusis and L.~Rastelli, \emph{{Correlation functions in the CFT(d) / AdS(d+1) correspondence}}, \href{https://doi.org/10.1016/S0550-3213(99)00053-X}{\emph{Nucl. Phys.} {\bfseries B546} (1999) 96} [\href{https://arxiv.org/abs/hep-th/9804058}{{\ttfamily hep-th/9804058}}].

\bibitem{Allen:1985ux}
B.~Allen, \emph{{Vacuum States in de Sitter Space}}, \href{https://doi.org/10.1103/PhysRevD.32.3136}{\emph{Phys. Rev.} {\bfseries D32} (1985) 3136}.

\bibitem{Dolan:2000ut}
F.~A. Dolan and H.~Osborn, \emph{{Conformal four point functions and the operator product expansion}}, \href{https://doi.org/10.1016/S0550-3213(01)00013-X}{\emph{Nucl. Phys.} {\bfseries B599} (2001) 459} [\href{https://arxiv.org/abs/hep-th/0011040}{{\ttfamily hep-th/0011040}}].

\bibitem{Dolan:2003hv}
F.~A. Dolan and H.~Osborn, \emph{{Conformal partial waves and the operator product expansion}}, \href{https://doi.org/10.1016/j.nuclphysb.2003.11.016}{\emph{Nucl. Phys.} {\bfseries B678} (2004) 491} [\href{https://arxiv.org/abs/hep-th/0309180}{{\ttfamily hep-th/0309180}}].

\bibitem{Dolan:2011dv}
F.~A. Dolan and H.~Osborn, \emph{{Conformal Partial Waves: Further Mathematical Results}},  \href{https://arxiv.org/abs/1108.6194}{{\ttfamily 1108.6194}}.

\bibitem{Leonhardt:2003qu}
T.~Leonhardt, R.~Manvelyan and W.~Ruhl, \emph{{The Group approach to AdS space propagators}}, \href{https://doi.org/10.1016/j.nuclphysb.2003.07.007}{\emph{Nucl. Phys.} {\bfseries B667} (2003) 413} [\href{https://arxiv.org/abs/hep-th/0305235}{{\ttfamily hep-th/0305235}}].

\bibitem{Liu:1998ty}
H.~Liu and A.~A. Tseytlin, \emph{{On four point functions in the CFT / AdS correspondence}}, \href{https://doi.org/10.1103/PhysRevD.59.086002}{\emph{Phys. Rev. D} {\bfseries 59} (1999) 086002} [\href{https://arxiv.org/abs/hep-th/9807097}{{\ttfamily hep-th/9807097}}].

\bibitem{Jorstad:2023ajr}
E.~J\o{}rstad, S.~Pasterski and A.~Sharma, \emph{{Equating extrapolate dictionaries for massless scattering}}, \href{https://doi.org/10.1007/JHEP02(2024)228}{\emph{JHEP} {\bfseries 02} (2024) 228} [\href{https://arxiv.org/abs/2310.02186}{{\ttfamily 2310.02186}}].

\bibitem{Furugori:2023hgv}
H.~Furugori, N.~Ogawa, S.~Sugishita and T.~Waki, \emph{{Celestial two-point functions and rectified dictionary}}, \href{https://doi.org/10.1007/JHEP02(2024)063}{\emph{JHEP} {\bfseries 02} (2024) 063} [\href{https://arxiv.org/abs/2312.07057}{{\ttfamily 2312.07057}}].

\bibitem{Nguyen:2023miw}
K.~Nguyen, \emph{{Carrollian conformal correlators and massless scattering amplitudes}}, \href{https://doi.org/10.1007/JHEP01(2024)076}{\emph{JHEP} {\bfseries 01} (2024) 076} [\href{https://arxiv.org/abs/2311.09869}{{\ttfamily 2311.09869}}].

\bibitem{Heemskerk:2009pn}
I.~Heemskerk, J.~Penedones, J.~Polchinski and J.~Sully, \emph{{Holography from Conformal Field Theory}}, \href{https://doi.org/10.1088/1126-6708/2009/10/079}{\emph{JHEP} {\bfseries 10} (2009) 079} [\href{https://arxiv.org/abs/0907.0151}{{\ttfamily 0907.0151}}].

\bibitem{El-Showk:2011yvt}
S.~El-Showk and K.~Papadodimas, \emph{{Emergent Spacetime and Holographic CFTs}}, \href{https://doi.org/10.1007/JHEP10(2012)106}{\emph{JHEP} {\bfseries 10} (2012) 106} [\href{https://arxiv.org/abs/1101.4163}{{\ttfamily 1101.4163}}].

\bibitem{Bekaert:2015tva}
X.~Bekaert, J.~Erdmenger, D.~Ponomarev and C.~Sleight, \emph{{Quartic AdS Interactions in Higher-Spin Gravity from Conformal Field Theory}}, \href{https://doi.org/10.1007/JHEP11(2015)149}{\emph{JHEP} {\bfseries 11} (2015) 149} [\href{https://arxiv.org/abs/1508.04292}{{\ttfamily 1508.04292}}].

\bibitem{Meltzer:2019nbs}
D.~Meltzer, E.~Perlmutter and A.~Sivaramakrishnan, \emph{{Unitarity Methods in AdS/CFT}}, \href{https://doi.org/10.1007/JHEP03(2020)061}{\emph{JHEP} {\bfseries 03} (2020) 061} [\href{https://arxiv.org/abs/1912.09521}{{\ttfamily 1912.09521}}].

\bibitem{Ferrara:1974ny}
S.~Ferrara, R.~Gatto and A.~F. Grillo, \emph{{Properties of Partial Wave Amplitudes in Conformal Invariant Field Theories}}, \href{https://doi.org/10.1007/BF02769009}{\emph{Nuovo Cim. A} {\bfseries 26} (1975) 226}.

\bibitem{Paulos:2016fap}
M.~F. Paulos, J.~Penedones, J.~Toledo, B.~C. van Rees and P.~Vieira, \emph{{The S-matrix bootstrap. Part I: QFT in AdS}}, \href{https://doi.org/10.1007/JHEP11(2017)133}{\emph{JHEP} {\bfseries 11} (2017) 133} [\href{https://arxiv.org/abs/1607.06109}{{\ttfamily 1607.06109}}].

\bibitem{Sleight:2021iix}
C.~Sleight and M.~Taronna, \emph{{On the consistency of (partially-)massless matter couplings in de Sitter space}}, \href{https://doi.org/10.1007/JHEP10(2021)156}{\emph{JHEP} {\bfseries 10} (2021) 156} [\href{https://arxiv.org/abs/2106.00366}{{\ttfamily 2106.00366}}].

\bibitem{Mack:2009gy}
G.~Mack, \emph{{D-dimensional Conformal Field Theories with anomalous dimensions as Dual Resonance Models}}, {\emph{Bulg. J. Phys.} {\bfseries 36} (2009) 214} [\href{https://arxiv.org/abs/0909.1024}{{\ttfamily 0909.1024}}].

\end{thebibliography}\endgroup

\end{document}